\documentclass[journal,10pt]{IEEEtran}

\usepackage{stfloats}
\usepackage{amsmath}
\usepackage{amssymb,latexsym}
\usepackage{graphicx}
\usepackage{color,cite,times}
\usepackage{epstopdf}
\usepackage{algorithmic}
\usepackage[ruled,linesnumbered] {algorithm2e}
\usepackage{booktabs}
\usepackage{bbm} % the Letter hollowing
\usepackage{url}
\usepackage{fancyhdr}
\usepackage{cleveref}
\usepackage{verbatim}
\usepackage{multirow}
\usepackage{makecell}
\usepackage{graphicx}
\usepackage{indentfirst}%大括号编号
\usepackage{cases}
\newcommand{\RNum}[1]{\uppercase\expandafter{\romannumeral #1\relax}}
\usepackage{colortbl} %表格颜色填充
\usepackage{extarrows}
\usepackage{threeparttable}
\usepackage{diagbox}
\usepackage{pifont}
\usepackage{cancel}
\usepackage{framed}
\usepackage{overpic}

\usepackage{subfigure}

\newcommand{\bsm}{\boldsymbol}

\newcommand{\mbf}{\mathbf}
\newcommand{\diag}{\mathrm{diag}}

\newtheorem{lemm}{Lemma}
\newtheorem{remk}{Remark}

\newtheorem{prop}{Proposition}

\definecolor{gray}{RGB}{192,192,192}

\begin{document}
\title{Multi-Device Task-Oriented Communication via Maximal Coding Rate Reduction}

\author{Chang Cai, \IEEEmembership{Graduate Student Member, IEEE,}
	Xiaojun Yuan, \IEEEmembership{Senior Member, IEEE,} \\
	and Ying-Jun Angela Zhang, \IEEEmembership{Fellow, IEEE} 
		\thanks{Chang Cai and Ying-Jun Angela Zhang are with the Department of Information Engineering, The Chinese University of Hong Kong, Hong Kong (e-mail: cc021@ie.cuhk.edu.hk; yjzhang@ie.cuhk.edu.hk).
			
			Xiaojun Yuan is with the National Key Laboratory of Wireless Communications,
			%the Center for Intelligent Networking and Communications, 
%				the National Key Laboratory of Science and Technology on Communications, 
			University of Electronic Science and Technology of China, Chengdu 611731, China (e-mail: xjyuan@uestc.edu.cn). 
			
			Code will be available at https://github.com/chang-cai/TaskCommMCR2 upon the acceptance of this paper.
		}
	}
\maketitle

\begin{abstract}
%	Task-oriented communication, which extracts only task-relevant information for transmission, is envisioned to be a key enabler to alleviate the communication burden in next-generation wireless networks.
	In task-oriented communications, most existing work designed the physical-layer communication modules and learning based codecs with distinct objectives: learning is targeted at accurate execution of specific tasks, while communication aims at optimizing conventional communication metrics, such as throughput maximization, delay minimization, or bit error rate minimization.
	The inconsistency between the design objectives may hinder the exploitation of the full benefits of task-oriented communications.
	In this paper, we consider a task-oriented multi-device edge inference system over a multiple-input multiple-output (MIMO) multiple-access channel, where the learning (i.e., feature encoding and classification) and communication (i.e., precoding) modules are designed with the same goal of inference accuracy maximization.
%	In this paper, we consider a specific task-oriented communication system for multi-device edge inference over a multiple-input multiple-output (MIMO) multiple-access channel, where the learning (i.e., feature encoding and classification) and communication (i.e., precoding) modules are designed with the same goal of inference accuracy maximization.
	Instead of end-to-end learning which involves both the task dataset and wireless channel during training, we advocate a separate design of learning and communication to achieve the consistent goal.
	Specifically, we leverage the maximal coding rate reduction (MCR$^2$) objective as a surrogate to represent the inference accuracy, which allows us to explicitly formulate the precoding optimization problem. 
	We cast valuable insights into this formulation and develop a block coordinate ascent (BCA) algorithm for efficient problem-solving.
%	Moreover, the MCR$^2$ objective also serves the loss function of the feature encoding network,
    Moreover, the MCR$^2$ objective serves the loss function for feature encoding and guides the classification design.
%	based on which we characterize the received features as a Gaussian mixture (GM) model, facilitating a maximum \textit{a posteriori} (MAP) classifier to infer the result.
	Simulation results on the synthetic features explain the mechanism of MCR$^2$ precoding at different SNRs. % mechanism
	We also validate on the CIFAR-10 and ModelNet10 datasets that the proposed design achieves a better latency-accuracy tradeoff compared to various baselines with inconsistent learning-communication objectives. %relying on the reconstruction of transmitted features.
%	Simulation results on both the synthetic and real-world datasets demonstrate the superior performance of the proposed method compared to various baselines.
	As such, our work paves the way for further exploration into the synergistic alignment of learning and communication objectives in task-oriented communication systems.
\end{abstract}

\begin{IEEEkeywords}
Task-oriented communication, maximal coding rate reduction (MCR$^2$), multi-device edge inference
\end{IEEEkeywords}

\section{Introduction}
As we step into the era of artificial intelligence (AI), the next-generation wireless networks are foreseeable to support various new usage scenarios such as autonomous vehicular networks, brain-computer interfaces, and smart healthcare \cite{6Groadmap, walid20206G, shi2022edgeAI}.
In such applications, the ultimate goal of communication is usually no longer the exact recovery of the underlying raw data, but the efficient execution of a certain task.
These applications typically generate unprecedented amounts of data (e.g., 10 Gb/s/m$^{\text{3}}$ traffic density) to serve various tasks,
in which the raw data exchange may pose an unaffordable burden on future communication systems.
To address this issue, task-oriented communications \cite{gunduz2023beyond}, a.k.a. semantic communications \cite{kaibin2021jcin, dusit2023survey}, are envisioned to be a key enabler to alleviate the communication burden.
In contrast to conventional communications that aim to recover the raw input data at the receiver, task-oriented communications focus on transmitting only the information pertinent to the task's objective, offering ample opportunities for reducing communication costs.

Thanks to the recent advances in AI, deep neural networks (DNNs) have been widely introduced in task-oriented communications to extract task-relevant features and serve various downstream tasks \cite{shao2020iccwksp, jiawei2021task, jiawei2023multidevice, djscc2019tccn, wir2021jsac, speech2021qin, zhijin2021single, zhijin2021multiuser, zhang2022unified, multimodal2022wcm, du2023jsac}.
In particular, the authors in \cite{shao2020iccwksp} proposed an end-to-end architecture to learn the on-device feature encoder and the server-side decoder for a classification task.
%The proposed network is directly trained with the cross-entropy (CE) loss. % for accurate classification.
The information bottleneck principle is investigated in \cite{jiawei2021task} to support variable-length feature encoding by imposing a sparsity-inducing prior on the learned features.
For image retrieval, the authors in \cite{djscc2019tccn, wir2021jsac} proposed a deep joint source and channel coding (JSCC) scheme based on the autoencoder architecture, where the wireless channel is incorporated in the autoencoder as a non-trainable layer.
An attention-based semantic communication system for speech transmission is developed in \cite{speech2021qin}.
More recently, there is an increasing interest in the study of serving different tasks with multimodal data simultaneously \cite{zhijin2021multiuser, zhang2022unified, multimodal2022wcm}.
%For example, the authors in \cite{zhijin2021multiuser} proposed a transformer based framework that shares the same transmitter structure for image retrieval, machine translation, and visual question answering (VQA) tasks.
%In \cite{zhang2022unified}, a multi-exit architecture is developed so that the decoder can provide early-exit results for relatively simple tasks.

The above work mainly focuses on the DNN-based codec design for different tasks, whereas the physical-layer design in task-oriented communications remains in the infancy stage. % did not rely on the considered task. % remains in the infancy stage
In \cite{jiawei2023multidevice}, the authors oversimplify the wireless transmission as error-free bit pipes, ignoring the interplay between wireless communications and learning tasks.
The intrinsic characteristics of wireless fading channel have been considered in \cite{zhijin2021single, zhijin2021multiuser, speech2021qin, zhang2022unified, multimodal2022wcm}.
Nonetheless, the physical-layer design criteria therein are still throughput maximization, delay minimization, or bit error rate (BER) minimization as in conventional communications, which are not aligned with the design objective of the learning module targeted at accurate execution of specific tasks.
For example, in \cite{zhijin2021single}, the mutual information serves as the loss function to train the transceivers for natural language processing (NLP) tasks.
Ref. \cite{zhijin2021multiuser} directly applies the linear minimum mean-square error (LMMSE) detector for exact recovery of transmitted features.
The inconsistency between learning and communication objectives may hinder the exploitation of the full benefits of task-oriented communications.
%Therefore, it is desirable to keep a common objective for learning and communication focusing on the successful completion of the task.
%Therefore, it is desirable to keep a common objective for learning and communication, i.e., the successful completion of the task.

%End-to-end learning, which parameterizes the learning and communication modules all by neural networks (NNs) and trains them in an end-to-end manner, can be a potential candidate to achieve a consistent design objective for learning and communication.
End-to-end learning, which parameterizes the learning and communication modules all by neural networks (NNs) and trains them in an end-to-end manner, can be a potential candidate to achieve a consistent design objective for learning and communication targeted at the successful completion of the task.
However, it is typically unaffordable to train such an end-to-end network adaptive to varying wireless environments, especially for multi-antenna transceivers that result in a high-dimensional channel matrix.
This is because the end-to-end network needs to learn the parameters based on both the task dataset and the entire distribution of wireless channels, incurring a prohibitively large training overhead and unpredictable training complexity \cite{wsc2023wcm}.
To resolve this issue, we advocate the separation of learning task and communication design, while maintaining a consistent design objective for both modules considering the successful execution of the task.
In this way, the training of the codecs does not involve the physical-layer wireless channel.
The state-of-the-art NN architectures in machine learning literature can be directly integrated into our framework.
On the other hand, the design of physical-layer modules does not require the individual training samples in the task dataset.
Instead, it only requires the channel state information and (possibly) some statistics of the learned features.
%Instead, it only requires the wireless channel and the statistics of the learned features.

\begin{figure*}
	[t]
	\centering
	%	\vspace{-1em}
	\includegraphics[width=1.6\columnwidth]{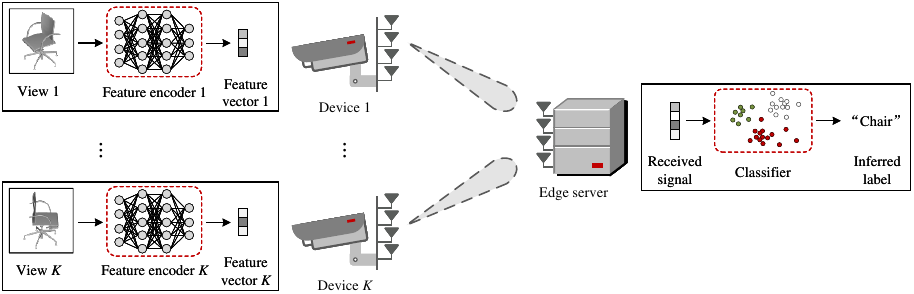}
	%	\vspace{-1em}
	\caption{An example of multi-device edge inference for multi-view object detection over a MIMO multiple-access channel.}
	\label{System_Model}
	%	\vspace{-1em}
\end{figure*}

%In this paper, we materialize the above design philosophy in a multi-device edge inference system performing a classification task.
In this paper, we materialize the above design philosophy in a task-oriented communication system for multi-device edge inference.
Compared to single-device edge inference \cite{shao2020iccwksp, jiawei2021task}, cooperation among multiple devices with distinct views improves the inference capability.
Multi-device edge inference has seen tremendous applications in pose estimation \cite{pose2021iccv}, 3D localization \cite{3Dlocalization2018tpami}, vehicle re-identification \cite{vehiclereid2018cvpr} and so on.
This paradigm has recently attracted much research attention in the design of wireless systems \cite{ensembles2021icassp, jiawei2023multidevice, collaborative2023wcl}.
However, as mentioned before, the communication design therein often directly reuses the metrics in traditional communications, which motivates the exploration on the consistent learning-communication objective targeted at inference accuracy maximization in this paper.

Fig. \ref{System_Model} shows an example of the wireless multi-device edge inference system performing a multi-view object detection task \cite{su2015multi}.
In the considered multi-device edge inference system,
the feature encoder at each device extracts low-dimensional features from its respective input data.
These extracted features undergo precoding before transmission over a multiple-input multiple-output (MIMO) multiple-access channel. % are precoded and transmitted
The edge server then processes the received features to output the inference result.
We aim to design the learning task (i.e., feature encoding and classification) and precoding optimization in a modularized way, both aiming at maximizing inference accuracy.
However, this objective introduces two main challenges.
The first challenge stems from the fact that the inference accuracy can only be empirically evaluated over all the testing samples.
It has no explicit expression as a function of the precoders and is thus hard to optimize directly.
The second challenge is more conceptual.
In our system, the edge server's goal is to output the inferred label, not to recover the transmitted features or the raw input data. 
Therefore, traditional signal detection is not essential for the considered inference task. 
However, it remains unclear how to execute classification directly based on the received features that are distorted by the wireless channel, without resorting to signal detection.

%To resolve the two aforementioned challenges, it is critical to explicitly characterize the classification accuracy based on the intermediate features.
%That is, to formulate a measure on the separability of the different classes of features as a surrogate of the classification accuracy.
To resolve the two aforementioned challenges, 
%To resolve the first challenge, 
%it is critical to explicitly characterize the inference accuracy relying on the intermediate features instead of the classifier's output, or,
%more precisely, to formulate a measure on the separability of the different classes of features as a surrogate of the inference accuracy.
it is critical to explicitly characterize the inference accuracy based on the intermediate features instead of relying on the classifier's output.
More precisely, we need to formulate a measure on the separability of the different classes of features as a surrogate of the inference accuracy.
This characterization/measure can serve as the loss function for feature encoding, directly promoting the desired behaviors for classification in the feature space.
More importantly, it can also be applied to measure the accuracy based on the received features, which is an explicit function of the precoders amenable to optimization, hence resolving the first challenge.
We note that the loss functions used in existing studies, such as cross-entropy (CE) \cite{zhijin2021single, zhijin2021multiuser, shao2020iccwksp, wir2021jsac} and information bottleneck \cite{jiawei2021task, jiawei2023multidevice}, rely heavily on the classifier's output, and thus cannot be applied to the precoding design problem that is separated from the learning task.
Regarding the second challenge, the proposed surrogate measure is expected to promote a distribution on the (received) features that is easy to compute.
Once this is achieved, we can implement a naive Bayes classifier to infer the result in place of the traditional detect-then-classify structure.

In this paper, we employ the principle of maximal coding rate reduction (MCR$^2$), as proposed in a recent study \cite{yu2020learning}, to tackle the two challenges simultaneously.
The MCR$^2$ is related to lossy data compression and is used as a tool to design and interpret modern deep networks \cite{chan2022redunet}.
It endeavors to expand the space spanned by the overall features and, in the meanwhile, to compress the subspace spanned by each individual class, so that different classes are maximally separated.
This work invokes the MCR$^2$ as a measure of inference accuracy for both feature encoding and precoding optimization design.
%This work invokes the MCR$^2$ as a measure of inference accuracy, which serves the objective both for feature encoding and for precoding optimization.
%To our best knowledge, this is the first work that introduces the principle of MCR$^2$ to communication field for physical-layer design.
% and can serve as a measure on the separability of different classes of the intermediate features.
%We invoke this objective both for feature encoding and for precoding optimization.
%Further, we characterize the distribution of the received features, enabling a channel-adaptive maximum \textit{a posteriori} (MAP) classifier to generate the inference result.
The main contributions of this paper are summarized as follows.
%applied

%the feature space behavior for the classification problem.
%More specifically, we endeavor to seek

%The characterization should describe 
%, and further, establish a measure of the inference accuracy based on the intermediate features.

%That is, to answer the following question:

%Based on the characterization, we are able to control the inference accuracy by shaping the feature space.

% 能够直接对feature 做loss 而不是依赖于label
% 需要一个目标函数 能够刻画feature space的behaviour
% 利用feature space的某种structure来做loss
% 把mcr2本身的motivation拿过来讲

%We tackle the two aforementioned challenges by employing the principle of maximal coding rate reduction (MCR$^2$), as proposed in a recent study \cite{yu2020learning}.
%%referring to a recent work \cite{yu2020learning} on the principle of maximal coding rate reduction (MCR$^2$).
%Coding rate reduction is related to lossy data compression and can serve as a measure on the separability of different classes of the intermediate features.
%We invoke this objective as a surrogate measure of the intractable inference accuracy for precoding optimization.
%Moreover, we adopt the MCR$^2$ as the loss function to train the feature encoders, thus unifying the objectives of learning and communication.
%We characterize the distribution of the received features, enabling a channel-adaptive maximum \textit{a posteriori} (MAP) classifier to generate the inference result.
%The main contributions of this paper are summarized as follows.

\begin{itemize}
	\item We formulate the precoding problem for inference accuracy maximization by maximizing the coding rate reduction objective measured on the received features, subject to the individual power constraints of devices.
	Intuitively, this formulation separates different classes of the received features as much as possible, hence improving the inference accuracy.
	
	\item To solve the problem, we derive an equivalent transform which non-trivially extends the idea behind the weighted minimum mean-square error (WMMSE) framework \cite{shi2011iteratively}.
	The equivalent problem allows for the use of a block coordinate ascent (BCA) solution algorithm, which is guaranteed to reach a stationary point of the original MCR$^2$ formulation. % algorithm to solve it efficiently
	
	\item In addition, we provide fundamental understanding on the MCR$^2$ formulation by analyzing the gradient of the precoder.
	Our analysis reveals that the precoding optimization aims to incrementally expand the volume spanned by the overall received features, and meanwhile compresses that spanned by each class of the received features.
	This process leads to a distribution with more separable classes at the server.
	
	\item We adopt the MCR$^2$ objective to learn the feature encoders, unifying the learning and communication objectives.
	We point out that the MCR$^2$ loss promotes a Gaussian mixture (GM) distribution on the learned features, with each class corresponding to a Gaussian component.
	After a linear transformation imposed by the wireless MIMO channel,
	the received features also follow a GM distribution, facilitating a channel-adaptive maximum \textit{a posteriori} (MAP) classifier to calculate the inference result.
\end{itemize}
Visualization results on the synthetic features reveal how the MCR$^2$ precoder reshapes the transmitted features.
%Visualization results on the synthetic features provides a better understanding on how the MCR$^2$ precoder reshapes the transmitted features.
We also evaluate the proposed design on the CIFAR-10 \cite{krizhevsky2009learning} and ModelNet10 \cite{wu20153d} datasets.
Our results demonstrate the superior performance of the proposed method compared to various baselines, especially when the communication resources (e.g., transmit power, time slots) are insufficient. % on the CIFAR-10 and ModelNet10 datasets 
%We evaluate our proposed method on both the synthetic features and real-world datasets.
%Our results demonstrate the superior performance of the proposed method compared to various baselines, especially when the communication resources (e.g., transmit power, time slots) are insufficient. % on the CIFAR-10 and ModelNet10 datasets 
%Moreover, the visualization on the synthetic features provides a better understanding on how the MCR$^2$ precoder reshapes the transmitted features.

The remainder of this paper is organized as follows.
In Section \ref{SMPD}, we describe the system model of multi-device edge inference and present general formulations of the learning and communication design problems.
In Section \ref{Section_MCR2}, we introduce the MCR$^2$ to specify the design tasks of the feature encoder, the precoder, and the classifier.
Section \ref{Section_Interpretation} gives valuable insights into the MCR$^2$ based precoding formulation.
In Section \ref{Section_Algorithm}, we propose an efficient algorithm to solve the precoding optimization problem.
Section \ref{Section_Simulation} provides extensive simulation results on different datasets. % both the synthetic features and real-world
Finally, we conclude this paper in Section \ref{Section_Conclusion}.

\textit{Notations:}
Lower-case letters are used to denote scalars.
Vectors and matrices are denoted by lower-case and upper-case boldface letters, respectively.
$\mbf{A}^{\sf T}$, $\mbf{A}^{\sf H}$, and $\mbf{A}^{-1}$ denote the transpose, conjugate transpose, and inverse of matrix $\mbf{A}$, respectively.
We use $\left[\mbf{a}\right]_{m:n}$ to denote the vector formed by the $m$-th to $n$-th entries of $\mbf{a}$.
We use $\jmath \triangleq \sqrt{-1}$, $\otimes$, $\diag\{ \cdot \}$, $\mathbb{E} ( \cdot )$, $\mathrm{tr} ( \cdot )$, and $\mathcal{R}(\cdot)$ to represent the imaginary unit, the Kronecker product, the diagonal operator, the expectation operator, the trace, and the range space, respectively.
The distribution of a complex Gaussian random vector with mean $\bsm{\mu}$, covariance matrix $\bsm{\Sigma}$, and relation matrix $\bsm{\Gamma}$ is denoted by $\mathcal{CN} (\bsm{\mu}, \bsm{\Sigma}, \bsm{\Gamma})$.
For a circularly symmetric complex Gaussian (CSCG) random vector with zero relation matrix, we denote its distribution as $\mathcal{CN} (\bsm{\mu}, \bsm{\Sigma})$.

\section{System Model and Problem Description} \label{SMPD}

\subsection{Multi-Device Edge Inference Model} \label{subsec_MDEI}
\begin{figure}
	[t]
	\centering
	%	\vspace{-1em}
	\includegraphics[width=1\columnwidth]{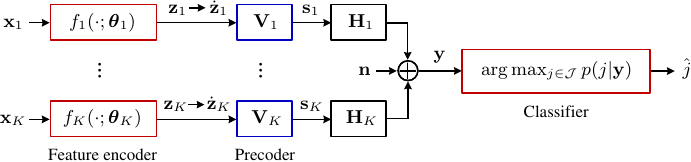} % diagram_0707.pdf
	%	\vspace{-1em}
	\caption{Flow diagram of the considered multi-device edge inference system.
		The red blocks correspond to the learning task for multi-device edge inference, and the blue blocks correspond to the communication design for precoding optimization.}
	\label{Diagram}
	\vspace{-1em}
\end{figure}

We consider a multi-device edge inference system, where $K$ edge devices collaborate with an edge server to perform an inference task, e.g., multi-view object detection \cite{su2015multi}, as shown in Fig. \ref{System_Model}.
Each edge device obtains its view of a common object from a particular perspective.
Denote by $\mbf{x}_k \in \mathbb{R}^{Q_k}$ the input view of device $k$, $k \in \mathcal{K} \triangleq \left\{1, \dots, K\right\}$.
%Let $\mbf{x} = \left[\mbf{x}_1^{\sf T}, \dots, \mbf{x}_K^{\sf T}\right]^{\sf T}$ be the input views of different edge devices.
Due to limited communication bandwidth, directly transmitting the raw data of each view to the edge server may cause significant delay.
Instead, each edge device, say device $k$, first extracts the feature $\mbf{z}_k \in \mathbb{R}^{D_k} (D_k \ll Q_k)$ from its input view $\mbf{x}_k$ via the feature encoder $f_k \left(\cdot ; \bsm{\theta}_k\right)$, which is realized by an NN with learnable parameters $\bsm{\theta}_k \in \mathbb{R}^{L_k}$.
Then, device $k$ transmits the precoded feature $\mbf{s}_k$ over the wireless channel.
Assume a total of $J$ classes, and define $\mathcal{J} \triangleq \left\{1, \dots, J\right\}$.
The edge server aims to infer the classification result $\hat{j} \in \mathcal{J}$ based on the received signal $\mbf{y}$.
%We elaborate the aforementioned steps in more detail as follows.
Fig. \ref{Diagram} presents the corresponding flow diagram.
The communication model is elaborated in the following subsection.

\subsection{Communication Model} \label{subsec_CM}
%\begin{figure}
%	[t]
%	\centering
%	%	\vspace{-1em}
%	\includegraphics[width=1\columnwidth]{Fig2_diagram.pdf} % diagram_0707.pdf
%	%	\vspace{-1em}
%	\caption{Flow diagram of the considered multi-device edge inference system.
%		The red blocks correspond to the learning task for multi-device edge inference, and the blue blocks correspond to the communication design for precoding optimization.}
%	\label{Diagram}
%	%	\vspace{-1em}
%\end{figure}
Consider linear analog modulation for feature transmission over a MIMO multiple-access channel. % uncoded 
Let $N_{{\mathsf t}, k}$ and $N_{\mathsf r}$ denote the number of antennas at device $k$ and at the edge server, respectively.
We reshape the real feature vector $\mbf{z}_k \in \mathbb{R}^{D_k}$ to the complex vector $\dot{\mbf{z}}_k \in \mathbb{C}^{\frac{D_k}{2}}$ to accommodate wireless transmission (assuming that $D_k$ is even).
In particular, the first and second halves of $\mbf{z}_k$ are mapped to the real and imaginary parts of $\dot{\mbf{z}}_k$, respectively, i.e.,
\begin{align} \label{real_complex}
	\dot{\mbf{z}}_k = \left[\mbf{z}_k\right]_{1:\frac{D_k}{2}} + \jmath \left[\mbf{z}_k\right]_{\frac{D_k}{2} +1 : D_k} .
\end{align}
Since the dimension of the complex feature vector $\dot{\mbf{z}}_k$ is generally greater than the effective channel rank, multiple time slots are needed to complete the transmission of one $\dot{\mbf{z}}_k$ for edge inference. %feature vector. %the number of transceiver antennas and
Denote by $T$ the number of required time slots, and define $\mathcal{T} \triangleq \left\{1, \dots, T\right\}$.

For device $k$, let $\mbf{V}_k (t) \in \mathbb{C}^{N_{{\sf t}, k} \times \frac{D_k}{2}}$ be the linear precoder at time slot $t$, $t \in \mathcal{T}$.
The transmitted signal vector $\mbf{s}_k (t) \in \mathbb{C}^{N_{{\sf t}, k}}$ at time slot $t$ is produced by
\begin{align}
	\mbf{s}_k(t) = \mbf{V}_k(t) \dot{\mbf{z}}_k. \label{precoding_t}
\end{align}
Let $\mbf{H}_k (t) \in \mathbb{C}^{N_{\mathsf r} \times N_{{\mathsf t}, k}}$ be the channel matrix from device $k$ to the edge server at time slot $t$, which is assumed known at the edge server.
The received signal vector $\mbf{y} (t) \in \mathbb{C}^{N_{\mathsf r}}$ at time slot $t$ is given by
\begin{align} \label{received_signal_t}
	\mbf{y} (t) = \sum_{k\in \mathcal{K} } \mbf{H}_k (t) \mbf{V}_k (t) \dot{\mbf{z}}_k + \mbf{n} (t),
\end{align}
where $\mbf{n} (t) \sim \mathcal{CN}\left(\mbf{0}, \delta_0^2\mbf{I}\right)$ is the independent CSCG noise vector at time slot $t$, with $\delta_0^2$ denoting the average noise power.

We concatenate the linear precoders at device $k$ for the total $T$ time slots by $\mbf{V}_k \triangleq \left[\mbf{V}_k^{\sf T}(1), \dots, \mbf{V}_k^{\sf T}(T) \right]^{\sf T} \in \mathbb{C}^{T N_{{\sf t}, k} \times \frac{D_k}{2}}$. % across
From \eqref{precoding_t}, at device $k$, the transmitted signal vector $\mbf{s}_k \triangleq \left[\mbf{s}_k^{\sf T} (1), \dots, \mbf{s}_k^{\sf T}(T)\right]^{\sf T} \in \mathbb{C}^{T N_{{\mathsf t}, k}}$ for the $T$ time slots is
\begin{align}
	\mbf{s}_k = \mbf{V}_k \dot{\mbf{z}}_k. \label{precoding}
\end{align}
The average transmit power at device $k$ is given by $\frac{1}{T}\mathbb{E} \left[ \left\|\mbf{s}_k\right\|_2^2 \right] = \frac{1}{T} \mathrm{tr} \left(\mbf{V}_k \mbf{\Sigma}^{(kk)} \mbf{V}_k^{\sf H}\right)$,
%\begin{align}
%	\frac{1}{T}\mathbb{E} \left[ \left\|\mbf{s}_k\right\|_2^2 \right] = \frac{1}{T} \mathrm{tr} \left(\mbf{V}_k \mbf{\Sigma}^{(kk)} \mbf{V}_k^{\sf H}\right),
%\end{align}
where $\mbf{\Sigma}^{(kk)} \in \mathbb{C}^{\frac{D_k}{2} \times \frac{D_k}{2}}$ is the covariance matrix of $\dot{\mbf{z}}_k$.
The feasible set of $\mbf{V}_k$ under the power constraint $P_k$ is expressed as 
%$\mathcal{V}_k \triangleq \left\{\mbf{V}_k \big| \mathrm{tr} \left(\mbf{V}_k \mbf{\Sigma}^{(kk)} \mbf{V}_k^{\sf H}\right) \leq T P_k\right\}$.
\begin{align} \label{feasible_set}
	\mathcal{V}_k \triangleq \left\{\mbf{V}_k \Big| \mathrm{tr} \left(\mbf{V}_k \mbf{\Sigma}^{(kk)} \mbf{V}_k^{\sf H}\right) \leq T P_k\right\}, ~~ k\in \mathcal{K}. 
\end{align}
Define $\mbf{H}_k \triangleq \diag \big\{\mbf{H}_k (1) ,\dots, \mbf{H}_k (T) \big\} \in \mathbb{C}^{T N_{\mathsf r} \times T N_{{\mathsf t}, k}}$.
The received signal vector for the $T$ time slots, denoted as $\mbf{y} \triangleq \left[\mbf{y}^{\sf T} (1), \dots, \mbf{y}^{\sf T}(T) \right]^{\sf T} \in \mathbb{C}^{TN_{\mathsf r}}$, is given by
\begin{align} \label{received_signal}
	\mbf{y} = \sum_{k\in \mathcal{K} } \mbf{H}_k \mbf{V}_k \dot{\mbf{z}}_k + \mbf{n},
\end{align}
where $\mbf{n} \triangleq \left[\mbf{n}^{\sf T} (1), \dots, \mbf{n}^{\sf T} (T)\right]^{\sf T}$ concatenates the CSCG noise vectors of the $T$ time slots.
Define $\mbf{H} \triangleq \left[\mbf{H}_1 , \dots, \mbf{H}_K\right] \in\mathbb{C}^{T N_{\mathsf r} \times T N_{{\mathsf t}}}$ with $N_{{\mathsf t}} \triangleq \sum_{k\in \mathcal{K}} N_{{\mathsf t}, k}$,
$\dot{\mbf{z}} \triangleq \left[\dot{\mbf{z}}_1^{\sf T}, \dots, \dot{\mbf{z}}_K^{\sf T}\right]^{\sf T} \in \mathbb{C}^{\frac{D}{2}}$ with $D \triangleq \sum_{k\in \mathcal{K}} D_k$,
and $\mbf{V} \triangleq \diag \left\{\mbf{V}_1, \dots, \mbf{V}_K\right\} \in \mathbb{C}^{T N_{{\mathsf t}} \times \frac{D}{2}}$.
We recast \eqref{received_signal} more compactly as
\begin{align} \label{received_signal_final}
	\mbf{y} = \mbf{H} \mbf{V} \dot{\mbf{z}}	+ \mbf{n}.
\end{align}

\subsection{Problem Description}
In this paper, we aim to design the feature encoder and the precoder at the edge devices, along with the classifier at the edge server, for the multi-device cooperative inference task. % with a unified goal of inference accuracy maximization.
%We list our design considerations as follows.
Our design considerations are described in the following.
\begin{itemize}
	\item \textit{Precoding optimization for inference accuracy maximization:}
	To fully unleash the potential of the task-oriented communication system, the criterion for precoding design should align with the learning task, which aims for inference accuracy maximization that deviates from the traditional focus on throughput maximization or MSE minimization \cite{zhijin2021multiuser} as in existing communication systems.
	If the feature encoder, the precoder, and the classifier are all parameterized by NNs, end-to-end learning of these modules could potentially maximize inference accuracy.
	However, in order to achieve good representation and generalization capabilities on the task dataset and over different channel realizations, the end-to-end network may be prohibitively large and challenging to train.
	Therefore, with the goal of inference accuracy maximization, we advocate separating the precoding design, which depends on the wireless channel, from the training process of the NN.

	\item \textit{Channel-adaptive classification bypassing signal detection:}
	Signal detection at the edge server to recover the transmitted feature $\dot{\mbf{z}}$ (or equivalently, $\mbf{z}$) is typically performed based on the MSE metric, which is not necessarily consistent with the learning task.
%	Thus, it may fail to exploit the full benefits of task-oriented communication.
	In fact, signal detection is not indispensable for the considered inference task. % indeed unnecessary
	In view of the above, we propose to bypass signal detection and infer the classification result directly based on the received signal $\mbf{y}$.
	Consequently, the classifier should be adaptive to the wireless channel.
	Implementing this adaptive classifier using a NN would be computationally infeasible due to the substantial training burden, considering both the task dataset and the variability of the wireless channel.
	Instead, we choose a naive Bayes classifier that calculates the probability of each class based on the probabilistic model of $\mbf{y}$.
	To facilitate this, the loss function for feature encoding is expected to promote a distribution on $\mbf{z}$ that is easy to compute.
%	To facilitate this, the loss function for training the feature encoders is expected to promote a distribution on $\mbf{z}$ that is easy to compute.
	Once this is achieved, the distribution of $\mbf{y}$ as a function of $\mbf{H}$ and $\mbf{V}$ can be obtained based on \eqref{received_signal_final} to facilitate channel-adaptive classification.
\end{itemize}

With the above considerations, the design problems of feature encoding, classification, and precoding optimization are stated as follows.
\subsubsection{Feature Encoding}
Denote the input data to the $K$ devices in concatenation by $\mbf{x} \triangleq \left[\mbf{x}_1^{\sf T}, \dots, \mbf{x}_K^{\sf T}\right]^{\sf T}$. % a concatenated form
Given a total of $M$ training data $\mbf{X} \triangleq \left[\mbf{x}^{(1)}, \dots, \mbf{x}^{(M)}\right]$ and the corresponding labels $\bsm{\Pi}$, % \triangleq \left[\bsm{\pi}^{(1)}, \dots, \bsm{\pi}^{(M)}\right]
the feature encoders are centrally trained to learn the parameters $\bsm{\theta} \triangleq \left\{\bsm{\theta}_k\right\}_{k \in \mathcal{K}}$ by minimizing the loss function $L \left(\mbf{X}, \mbf{\Pi}; \bsm{\theta}\right)$, i.e., %  to learn the feature mapping $\mbf{z} = f\left(\mbf{x}, \bsm{\theta}\right)$
\begin{align} \label{general_loss}
	\min_{\bsm{\theta}} \quad & L \left(\mbf{X}, \mbf{\Pi}; \bsm{\theta}\right).
\end{align}
The feature encoding problem is to design $L \left(\mbf{X}, \mbf{\Pi}; \bsm{\theta}\right)$ to extract features for accurate classification.
Meanwhile, it is expected to promote some easily computable distribution on the learned feature $\mbf{z}$.
We remark that the learning problem in \eqref{general_loss} does not involve the time-varying wireless channel.

\subsubsection{Classification}
We apply the MAP classifier at the edge server to infer the classification result:  %utilized to obtain the classification result:
\begin{align} \label{map}
	\hat{j} = \arg \max_{j \in \mathcal{J}} ~ p\left(j \big| \mbf{y}\right)
	= \arg \max_{j \in \mathcal{J}} ~ p_j p\left(\mbf{y} \big| j\right),
\end{align}
where $p_j$, $ p\left(j \big| \mbf{y}\right)$ and $p\left(\mbf{y} \big| j\right)$ denote the prior probability, the posterior probability and the likelihood of $\mbf{y}$, respectively.

\subsubsection{Precoding Optimization}
	We aim to maximize the inference accuracy $\mathrm{Acc} \left(\mbf{V}_1, \dots, \mbf{V}_K\right)$, subject to the average power constraint at each edge device.
	The optimization problem is formulated as
%	\begin{subequations} \label{geneal_precoding_opt}
%		\begin{align}
%			\max_{\left\{\mbf{V}_k\right\}_{k\in \mathcal{K}}} \quad & \mathrm{Acc} \left(\mbf{V}_1, \dots, \mbf{V}_K\right) \label{inference_acc} \\ 
%			\operatorname{ s.t. } ~~\quad
%			& \frac{1}{T}\mathrm{tr} \left(\mbf{V}_k\mbf{\Sigma}^{(kk)} \mbf{V}_k^{\sf H}\right) \leq P_k, ~~k \in \mathcal{K}. 
%		\end{align}
%	\end{subequations}
	\begin{align} \label{geneal_precoding_opt}
		\max_{\left\{\mbf{V}_k \in \mathcal{V}_k\right\}_{k\in \mathcal{K}}} ~ \mathrm{Acc} \left(\mbf{V}_1, \dots, \mbf{V}_K\right). 
	\end{align}
%	\color{blue}
%	It is expected that the above formulation does not require the individual samples in the task dataset.
%	\color{black}
	In general, $\mathrm{Acc} \left(\mbf{V}_1, \dots, \mbf{V}_K\right)$ has no closed-form expression as a function of the precoders. % cannot be expressed explicitly
	This motivates us to find a surrogate measure of this intractable inference accuracy.
%	In the next section, we introduce a measure that represents the inference accuracy based on the separability of different classes of intermediate features.
	In the next section, we represent the inference accuracy by a measure on the separability of different classes of intermediate features.
%	We then apply this measure to the received features as a surrogate of $\mathrm{Acc} \left(\mbf{V}_1, \dots, \mbf{V}_K\right)$.
%	Applying this measure to the received features yields an explicit function of the precoders amenable to optimization.
	This measure is then applied to the received features, resulting in an explicit expression of the precoders amenable to optimization.
	
%it is critical to explicitly characterize the inference accuracy based on the intermediate features instead of relying on the classifier's output.
%More precisely, we need to formulate a measure on the separability of the different classes of features as a surrogate of the inference accuracy.
%This characterization/measure can serve as the loss function for feature encoding, directly promoting the desired behaviors for classification in the feature space.
%More importantly, it can also be applied to measure the accuracy based on the received features, which is an explicit function of the precoders amenable to optimization, hence resolving the first challenge.

\section{MCR$^2$ Based Task-Oriented Communication System} \label{Section_MCR2}
We first present some preliminaries on the MCR$^2$.
Then, we materialize the proposed task-oriented communication system, where the feature encoding design in \eqref{general_loss} and the precoding optimization in \eqref{geneal_precoding_opt} are both formulated based on the MCR$^2$ objective.
\subsection{Preliminaries}
For classification tasks, it is desirable for features of different classes to be as separable (or discriminative) as possible, while features of the same class should be correlated and coherent.
In line with this spirit, the authors in \cite{yu2020learning} proposed to promote subspace structures on the features. % propose to promote advocate
In particular, features of different classes should lie in different subspaces which collectively span a space of the largest possible volume (or dimension).
Conversely, features of the same class should only span a low-dimensional subspace of a minimal volume.
The authors in \cite{yu2020learning} resorted to the coding rate \cite{ma2007segmentation}, a concept from lossy data compression, as a measure of volume. % characterize the volume spanned by the feature samples
Accordingly, the desired features should yield a large difference between the coding rate for encoding them as a whole and that for encoding different classes separately.
The difference is referred to as coding rate reduction, which is defined as follows. % mathematically 

\begin{figure}
	[t]
%	\color{blue}
	\centering
	%	\vspace{-1em}
	\includegraphics[width=.55\columnwidth]{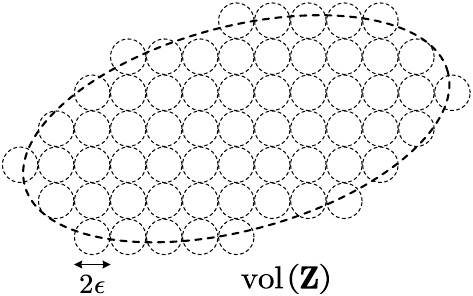}
	%	\vspace{-.5em}
	\caption{We pack the space spanned by the data $\mbf{Z}$ with small balls of radius $\epsilon$.
		The number of balls needed to pack the space gives the number of bits needed to record the location of each data point $\mbf{z}^{(m)}$ up to the given precision $\epsilon$.}
	\label{sphere_packing}
	%	\vspace{-.5em}
\end{figure}

\begin{figure}
	[t]
%	\color{blue}
	\centering
	%	\vspace{-1em}
	\includegraphics[width=1\columnwidth]{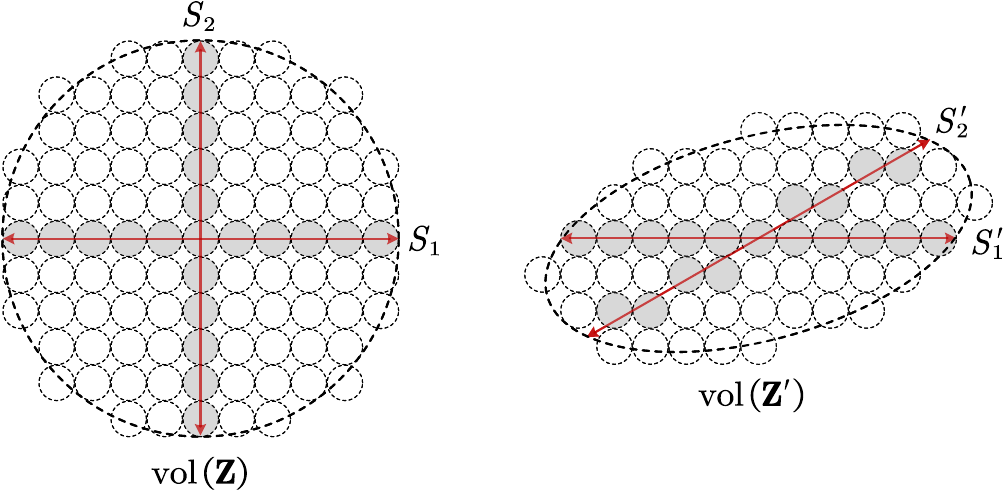}
	%	\vspace{-.5em}
	\caption{Comparison of two representations $\mbf{Z}$ and $\mbf{Z}^\prime$ via reduced rates, where $S_1 (S_1^\prime)$ and $S_2 (S_2^\prime)$ are the subspaces spanned by the first and second classes, respectively.
		In \eqref{coding_rate_reduction}, $R$ corresponds to the number of $\epsilon$-balls packed in the joint distribution (the total number of white and gray balls), $R_c$ corresponds to the number of gray balls, and $\Delta R$ corresponds to the number of white balls.
		The MCR$^2$ principle prefers the left representation.}
	\label{ccr_sphere_packing}
	%	\vspace{-.5em}
\end{figure}

Denote the features extracted at the $K$ devices in concatenation by $\mbf{z} \triangleq \left[\mbf{z}_1^{\sf T}, \dots, \mbf{z}_K^{\sf T}\right]^{\sf T}$.
Given $M$ feature samples $\mbf{Z} = \left[\mbf{z}^{(1)}, \dots, \mbf{z}^{(M)}\right] \in \mathbb{R}^{D \times M}$, the coding rate for encoding the samples up to the distortion $\epsilon$, i.e., $\mathbb{E} \left[\left\|\mbf{z} - \hat{\mbf{z}} \right\|_2^2 \right] \leq \epsilon^2$ for the decoded $\hat{\mbf{z}}$, is upper bounded by % allowable % as a whole % tightly
\begin{align} \label{coding_rate}
	R \left(\mbf{Z}, \epsilon\right) = \frac{1}{2} \log \det \left(\mbf{I} + \frac{D}{M \epsilon^2} \mbf{Z} \mbf{Z}^{\sf T}\right).
\end{align}
%\color{blue}
The coding rate function characterizes the average number of bits needed to encode each $\mbf{z}$.
This formula can be derived either by packing $\epsilon$-balls into the space spanned by $\mbf{Z}$ as a Gaussian source, or by computing the number of bits needed to quantize the singular value decomposition (SVD) of $\mbf{Z}$ subject to the precision $\epsilon$ \cite{ma2007segmentation}.
% therefore serving as a measure of volume.
See Fig. \ref{sphere_packing} for an illustration.
%\color{black}
%Assume a total of $J$ classes in $\mbf{Z}$, and define $\mathcal{J} \triangleq \left\{1, \dots, J\right\}$.
Let $\mbf{\Pi} = \left\{ \mbf{\Pi}_j \in \mathbb{R}^{M \times M} \right\}_{j\in \mathcal{J}}$ be a set of diagonal matrices, where the $m$-th diagonal entry of $\mbf{\Pi}_j$ denotes the probability of the $m$-th sample belonging to the $j$-th class, $m \in \mathcal{M} \triangleq \left\{1, \dots, M\right\}$, $j \in \mathcal{J}$, and we have $\sum_{j\in\mathcal{J}} \mbf{\Pi}_j = \mbf{I}$.
If we choose to encode the features of different classes separately, then the corresponding coding rate (subject to the allowable distortion $\epsilon$) is upper bounded by
\begin{align} \label{class_coding_rate}
	R_c\left(\mbf{Z}, \epsilon | \mbf{\Pi}\right) = \sum_{j\in \mathcal{J}} \frac{\mathrm{tr}\left(\mbf{\Pi}_j\right)}{2M} \log \det \left(\mbf{I} + \frac{D}{\mathrm{tr}\left(\mbf{\Pi}_j\right) \epsilon^2} \mbf{Z} \mbf{\Pi}_j \mbf{Z}^{\sf T}\right).
\end{align}
Consequently,
the coding rate reduction is given by %With $\mbf{\Pi}$ given
\begin{align} \label{coding_rate_reduction}
	\Delta R \left(\mbf{Z}, \mbf{\Pi}, \epsilon\right) = R \left(\mbf{Z}, \epsilon\right)
	- R_c\left(\mbf{Z}, \epsilon | \mbf{\Pi}\right).
\end{align}
%\color{blue}
Fig. \ref{ccr_sphere_packing} provides an interpretation to \eqref{coding_rate_reduction} from the sphere packing perspective.
%\color{black}

\begin{figure}
	[t]
	\centering
	%	\vspace{-1em}
	\includegraphics[width=.85\columnwidth]{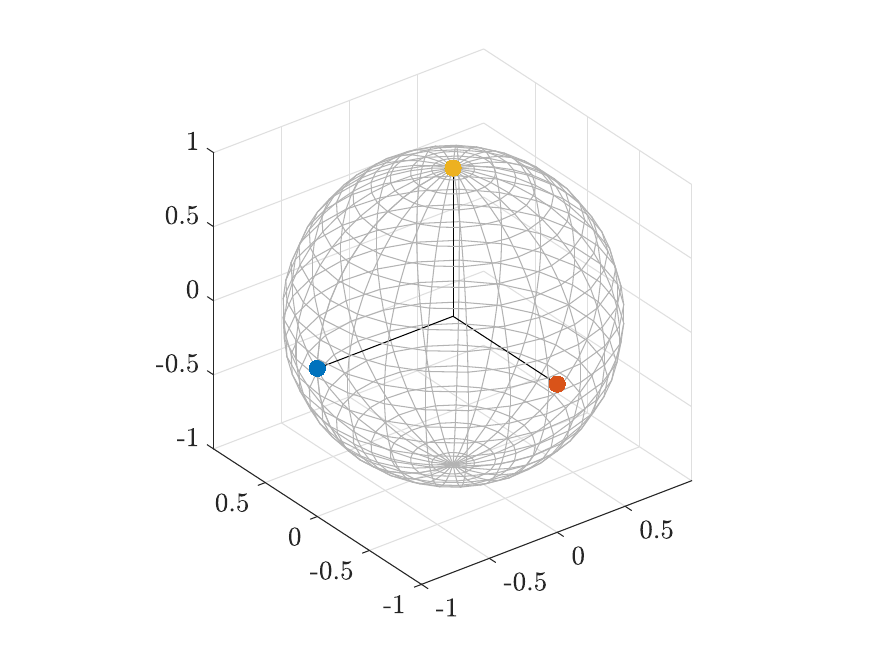}
%	\vspace{-.5em}
	\caption{Illustration of the optimal solution to the problem in \eqref{mcr2}.
		Assume $\mbf{z} \in \mathbb{R}^3$ with a total of $3$ classes, and the rank of each class is $1$. % of features
		The orange, blue, and yellow points represent the three different classes of features.
		The angle between any two different classes of features is $\pi / 2$.}
	\label{orthogonal_subspaces}
%	\vspace{-.5em}
\end{figure}

We aim to find the representation $\mbf{Z}$ that maximizes the coding rate reduction objective, a principle referred to as the maximal coding rate reduction (MCR$^2$).
Since $\Delta R \left(\mbf{Z}, \mbf{\Pi}, \epsilon\right)$ monotonically increases with the scale of $\mbf{Z}$, we normalize each feature sample on the unit ball such that different representations can be compared fairly.
The optimization problem is formulated as
\begin{subequations} \label{mcr2}
	\begin{align}
		\max_{\mbf{Z}} \quad & \Delta R \left(\mbf{Z}, \mbf{\Pi}, \epsilon\right)  \\ 
		\operatorname{ s.t. } \quad
		& \left\|\mbf{z}^{(m)} \right\|_2^2 = 1, ~~m \in \mathcal{M}. \label{sphere}
	\end{align}
\end{subequations}
Under the conditions of large feature dimension and high coding precision, the optimal solution to \eqref{mcr2} ensures that the features of different classes lie in distinct orthogonal subspaces. %are all orthogonal to each other. % of \eqref{mcr2}
% Fig. \ref{orthogonal_subspaces} for an illustration
Please see Fig. \ref{orthogonal_subspaces} for a visual illustration.
For a more rigorous explanation of this statement, please see \cite[Theorem 12]{chan2022redunet}.

\subsection{MCR$^2$ for Feature Encoding and Classification} \label{mcr2_4_ei}
Notably, the upper bound in \eqref{class_coding_rate} is achieved when each class of the feature samples follows a respective Gaussian distribution \cite{ma2007segmentation}.
In this case, the feature samples together are GM distributed, and \eqref{coding_rate} serves as a tight upper bound of the coding rate for encoding them as a whole.
From the above, the MCR$^2$ criterion actually promotes a GM distribution on the feature samples $\mbf{z}$.
We thus propose to train the distributed feature encoder by the MCR$^2$ loss function to learn $\bsm{\theta}$, i.e., % the feature mapping $\mbf{z} = f\left(\mbf{x}; \bsm{\theta}\right)$   %  that promotes a GM distribution on $\mbf{z}$
\begin{subequations} \label{mcr2_learning}
	\begin{align}
		\max_{\bsm{\theta}} \quad & \Delta R \left(\mbf{Z} \left(\bsm{\theta}\right), \mbf{\Pi}, \epsilon\right)  \\ 
		\operatorname{ s.t. } \quad
		& \left\|\mbf{z}^{(m)} \left(\bsm{\theta}\right)\right\|_2^2 = 1, ~~m \in \mathcal{M}. 
	\end{align}
\end{subequations}
It is noteworthy that 
even though the conditions in \cite[Theorem 12]{yu2020learning} may not hold for the above learning problem, empirical results show that the learned features of different classes tend to be maximally separated to each other.

Since $\mbf{z}$ is modeled as a real GM random vector, $\dot{\mbf{z}}$ is a complex GM random vector according to \eqref{real_complex}. 
The probability density function (PDF) of $\dot{\mbf{z}}$ is given by
\begin{align} \label{gm_pdf}
	p\left(\dot{\mbf{z}}\right) = \sum_{j \in \mathcal{J}} p_j \mathcal{CN} \left(\dot{\mbf{z}} ; \bsm{\mu}_j, \mbf{\Sigma}_j, \mbf{\Gamma}_j\right) ,
\end{align}
where $\bsm{\mu}_j$, $\mbf{\Sigma}_j$, and $\mbf{\Gamma}_j$ denote the mean vector, covariance matrix, and relation matrix of $\dot{\mbf{z}}$ in class $j$, respectively.
From \eqref{gm_pdf}, we have $p\left(\dot{\mbf{z}} \big| j\right) = \mathcal{CN}\left(\dot{\mbf{z}}; \bsm{\mu}_j, \mbf{\Sigma}_j, \mbf{\Gamma}_j \right)$.
Then, according to \eqref{received_signal}, the likelihood $p\left(\mbf{y} \big| j\right)$ is expressed as
\begin{align}
	p\left(\mbf{y} \big| j\right) &= \mathcal{CN} \big(\mbf{y}; \mbf{H}\mbf{V} \bsm{\mu}_j,\nonumber \\
	 & ~~~~~~~~~\mbf{H}\mbf{V} \mbf{\Sigma}_j \mbf{V}^{\sf H} \mbf{H}^{\sf H} + \delta_0^2 \mbf{I}, \mbf{H}\mbf{V} \mbf{\Gamma}_j \mbf{V}^{\sf T} \mbf{H}^{\sf T} \big).
\end{align}
Therefore, the MAP classifier in \eqref{map} is specified as
\begin{align} \label{map-specify}
	\hat{j} &= \arg \max_{j \in \mathcal{J}} ~ p_j \mathcal{CN} \big(\mbf{y}; \mbf{H}\mbf{V} \bsm{\mu}_j,  \nonumber \\
	 &~~~~~~~~~~~~~~~~~  \mbf{H}\mbf{V} \mbf{\Sigma}_j \mbf{V}^{\sf H} \mbf{H}^{\sf H} + \delta_0^2 \mbf{I}, \mbf{H}\mbf{V} \mbf{\Gamma}_j \mbf{V}^{\sf T} \mbf{H}^{\sf T} \big).
\end{align}

\subsection{MCR$^2$ for Precoding Optimization} \label{mcr2_4_precoding} % Problem Formulation
Since the classifier takes the received signal $\mbf{y}$ as input, the inference accuracy essentially depends on how separable (or discriminative) different classes of $\mbf{y}$ are.
We invoke the coding rate reduction objective measured on $\mbf{y}$ as a surrogate of the intractable inference accuracy, which can be expressed as a function of the precoders as
\begin{align}
	& \Delta R \left(\mbf{V}_1, \dots, \mbf{V}_K\right) \nonumber \\
	=\, & \log \det \left(\mbf{I} + \alpha \left(\mbf{H}\mbf{V} \mbf{\Sigma} \mbf{V}^{\sf H} \mbf{H}^{\sf H} + \delta_0^2 \mbf{I} \right) \right)  \nonumber \\
	&-\sum_{j \in \mathcal{J}} p_j
	\log \det \left(\mbf{I} + \alpha \left(\mbf{H}\mbf{V} \mbf{\Sigma}_j \mbf{V}^{\sf H} \mbf{H}^{\sf H} +  \delta_0^2 \mbf{I} \right) \right) \\
	=\,&  \underbrace{ \log \det \left(\gamma\mbf{I} + \alpha \mbf{H}\mbf{V} \mbf{\Sigma} \mbf{V}^{\sf H} \mbf{H}^{\sf H} \right)}_{R \left(\mbf{V}_1, \dots, \mbf{V}_K\right)} \nonumber \\
	&- \underbrace{\sum_{j \in \mathcal{J}} p_j
		\log \det \left( \gamma\mbf{I} + \alpha \mbf{H}\mbf{V} \mbf{\Sigma}_j \mbf{V}^{\sf H} \mbf{H}^{\sf H}  \right)}_{R_c \left(\mbf{V}_1, \dots, \mbf{V}_K\right)}, \label{mcr2-precoding}
\end{align}
where $\mbf{\Sigma}$ is the covariance matrix of $\dot{\mbf{z}}$, $\alpha \triangleq \frac{T N_{\sf r}}{ \epsilon^2}$, and $\gamma = 1+\alpha \delta_0^2$.
Notably, \eqref{mcr2-precoding} evaluates the inference accuracy without the information of individual samples in the task dataset.
Instead, it relies on the covariance matrix of the transmitted features and that in different classes.
The problem in \eqref{geneal_precoding_opt} is rewritten as
\begin{align}
	\text{(P1)}~\max_{\left\{\mbf{V}_k \in \mathcal{V}_k\right\}_{k\in \mathcal{K}}} ~ \Delta R \left(\mbf{V}_1, \dots, \mbf{V}_K\right). \label{mcr2-obj}
\end{align}
%\begin{subequations}
%	\begin{align}
%		\text{(P1)}~\max_{\left\{\mbf{V}_k\right\}_{k\in \mathcal{K}}} \quad & \Delta R \left(\mbf{V}_1, \dots, \mbf{V}_K\right) \label{mcr2-obj} \\ 
%		\operatorname{ s.t. }~~ \quad
%		& \frac{1}{T} \mathrm{tr}\left(\mbf{V}_k\mbf{\Sigma}^{(kk)} \mbf{V}_k^{\sf H}\right) \leq P_k, ~~k \in \mathcal{K} \label{power_constraint}.
%	\end{align}
%\end{subequations}
The primary challenge in solving (P1) lies in the non-convex MCR$^2$ objective, which involves the difference of log-determinant terms.
Moreover, the precoder at each device are strongly coupled in $\Delta R \left(\mbf{V}_1, \dots, \mbf{V}_K\right)$.
We emphasize on the importance of precoding optimization for the following reasons.
The learning objective in \eqref{mcr2_learning} for feature encoding guarantees maximally separated $\mbf{z}$ for different classes.
However, due to the linear transformation by the wireless MIMO channel,
the received signal $\mbf{y}$ may not maintain the same level of separability as $\mbf{z}$, %$\mbf{H}$ (or equivalently, $\bar{\mbf{H}}$) %  with noises added
leading to a deteriorated inference accuracy.
In (P1), the precoders are expected to promote the separability of different classes of $\mbf{y}$ by compensating for the channel distortion.
%Before digging into the algorithm design for solving (P1), we provide some insights into the MCR$^2$ guided precoding optimization in the next section. %in the following subsection

\subsection{Overall Implementation Procedures}
We summarize the implementation procedures of the MCR$^{\text{2}}$ based design as follows.
Firstly, the feature encoding network is centrally trained on the task dataset before deployment, as described in Section \ref{mcr2_4_ei}.
Then, based on the trained model, the feature statistics $\mbf{\Sigma}$ and $\left\{\bsm{\mu}_j, \mbf{\Sigma}_j, \mbf{\Gamma}_j\right\}_{j \in \mathcal{J}}$ are calculated by averaging the feature samples of the training data.
These statistics are then used for precoding optimization and classification during the inference stage, as shown in \eqref{mcr2-obj} and \eqref{map-specify}, respectively.
%Before digging into the algorithm design for solving (P1), we provide some insights into the MCR$^2$ guided precoding optimization in the next section. %in the following subsection
%The precoding optimization algorithm is elaborated in Section \ref{Section_Algorithm}.

\section{Interpretation of Precoding Optimization via Gradient Ascent} \label{Section_Interpretation}
Before digging into the algorithm design for solving (P1), we provide some insights into the MCR$^2$ guided precoding optimization in this section. 
Specifically, we are interested in how the precoding optimization reshapes the received signals to achieve a better inference accuracy. % for better classification
We consider the single-device case for ease of exposition. % (by dropping the user index)
We drop the user index and, with a slight abuse of notation, denote by $\mbf{V}$ the precoder at the device.
Given an initial point of $\mbf{V}$, a straightforward way to solve (P1) is to update $\mbf{V}$ iteratively via gradient based methods.
This motivates us to examine how the space spanned by the received noise-free features will change by following the gradient ascent of $\mbf{V}$ at each iteration.\footnote{We do not take the channel noise into account since precoding optimization cannot reshape the noise distribution.} % i.e., $\mathcal{R} \left(\mbf{H} \mbf{V} \mbf{\Sigma}^{\frac{1}{2}}\right)$ be updated
Notice that although updating $\mbf{V}$ strictly along its gradient direction may not be feasible due to the average power constraint, the gradient direction can still provide some valuable insights into the updating tendency. % remains to

The gradient matrix of $\Delta R \left(\mbf{V}\right)$ is expressed as
\begin{align} \label{gradient}
	\nabla \left(\Delta R \left(\mbf{V}\right)\right) =  \nabla R\left(\mbf{V}\right) -  \nabla R_c\left(\mbf{V}\right),
\end{align}
where
\begin{align}
	\nabla R\left(\mbf{V}\right)
	&= 2\alpha \mbf{H}^{\sf H} \mbf{D} \mbf{H} \mbf{V} \mbf{\Sigma} , \label{gradient_1}\\
	\nabla R_c\left(\mbf{V}\right)
	&= 2\alpha \sum_{j \in \mathcal{J}} p_j \mbf{H}^{\sf H} \mbf{D}_j \mbf{H} \mbf{V} \mbf{\Sigma}_j, \label{gradient_2}
\end{align}
with $\mbf{D} \triangleq \left(\gamma\mbf{I} + \alpha \mbf{H}\mbf{V} \mbf{\Sigma} \mbf{V}^{\sf H} \mbf{H}^{\sf H}\right)^{-1}$ and $\mbf{D}_j \triangleq \left(\gamma\mbf{I} + \alpha \mbf{H}\mbf{V} \mbf{\Sigma}_j \mbf{V}^{\sf H} \mbf{H}^{\sf H}\right)^{-1}$.
For given $\mbf{V}^t$ at iteration $t$, the update rule for gradient ascent is
\begin{align} \label{gradient_ascent}
	\mbf{V}^{t+1} =\mbf{V}^t + \eta  \nabla \left(\Delta R \left(\mbf{V}^t\right)\right) ,
\end{align}
where $\eta$ is the step size.
As mentioned before, we are interested in the space spanned by the received noise-free features, i.e., $\mathcal{R} \big(\mbf{H} \mbf{V} \mbf{\Sigma}^{\frac{1}{2}}\big)$, where $\mbf{\Sigma}^{\frac{1}{2}}$ denotes the square root \cite{horn2012matrix} of the covariance matrix $\mbf{\Sigma}$.
According to \eqref{gradient_ascent}, the increment of $\mbf{H}\mbf{V}\mbf{\Sigma}^{\frac{1}{2}}$ between iteration $t$ and $t+1$ is expressed as
%Substituting \eqref{gradient_ascent} into $\mbf{H} \mbf{V}^{t+1} \mbf{\Sigma}^{\frac{1}{2}}$ yields
\begin{align} \label{increment}
	&\mbf{H} \mbf{V}^{t+1} \mbf{\Sigma}^{\frac{1}{2}} - \mbf{H} \mbf{V}^t  \mbf{\Sigma}^{\frac{1}{2}} \nonumber \\
	=\, & \eta   \mbf{H} \nabla R \left(\mbf{V}^t\right) \mbf{\Sigma}^{\frac{1}{2}} 
	- \eta  \mbf{H} \nabla R_c \left(\mbf{V}^t\right) \mbf{\Sigma}^{\frac{1}{2}}.
	%	+ \eta \cdot  \mbf{H} \nabla R \left(\mbf{V}^t\right) \mbf{C}^{\frac{1}{2}}
\end{align}

\begin{figure}
	[t]
	\centering
	%	\vspace{-1em}
	%	\includegraphics[width=.6\columnwidth]{GD.pdf}
	\begin{overpic}[width=.45\textwidth]{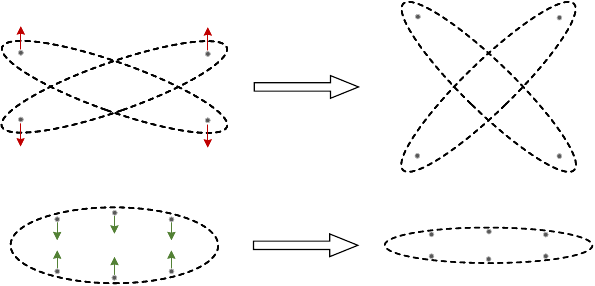}
		\put(48,35.8){\textcolor[RGB]{192,0,0}{\small$\mbf{E}^t$}}
		\put(45.6,9.8){\textcolor[RGB]{84,130,53}{\small$-\mbf{C}_j^t$}}
	\end{overpic}
	%	\vspace{-1em}
	\caption{Interpretation of precoding optimization with matrix operators.
		In the column space of $\mbf{P}$,
		$\mbf{E}^t$ expands the received features by contrasting them across different classes,
		while $- \mbf{C}^t_j$ compresses each class of the received features by contracting each class to a low-dimensional subspace.
		For clarity, the effect brought by the scaling factor $\mbf{\Lambda}^2$ is not sketched.}
	\label{GA}
	%	\vspace{-1em}
\end{figure}

The term $\eta   \mbf{H} \nabla R \left(\mbf{V}^t\right) \mbf{\Sigma}^{\frac{1}{2}}$ in \eqref{increment} corresponds to the increment brought by $\nabla R\left(\mbf{V}^t\right)$. %, i.e., the first term in \eqref{gradient}
With \eqref{gradient_1}, we obtain
%We express it explicitly as % the second term in \eqref{increment}
\begin{align} \label{increment_1st_term}
	\eta  \mbf{H} \nabla R \left(\mbf{V}^t\right) \mbf{\Sigma}^{\frac{1}{2}} 
	=2\eta \alpha \mbf{H} \mbf{H}^{\sf H} \mbf{D}^t \mbf{H} \mbf{V}^t  \mbf{\Sigma}^{\frac{3}{2}},
\end{align}
where $\mbf{D}^t \triangleq \left(\gamma\mbf{I} + \alpha \mbf{H}\mbf{V}^t \mbf{\Sigma} \left(\mbf{V}^t\right)^{\sf H} \mbf{H}^{\sf H}\right)^{-1}$ and $\mbf{\Sigma}^{\frac{3}{2}} \triangleq \mbf{\Sigma} \mbf{\Sigma}^{\frac{1}{2}}$.
Denote the SVD of $\mbf{H}$ as $\mbf{H} = \mbf{P} \mbf{\Lambda} \mbf{Q}^{\sf H}$.
It follows that $\mbf{H} \mbf{H}^{\sf H} = \mbf{P} \mbf{\Lambda}^2 \mbf{P}^{\sf H}$.
%The columns of $\mbf{P}$ form an orthonormal basis of $\mathcal{R}\left(\mbf{H}\right)$. % the column space of $\mbf{H}$.
Let $\mbf{G}^t \triangleq  \mbf{\Lambda} \mbf{Q}^{\sf H} \mbf{V}^t  \mbf{\Sigma}^{\frac{1}{2}}$ be the representation of $\mbf{H} \mbf{V}^t \mbf{\Sigma}^{\frac{1}{2}}$ under the orthonormal basis $\mbf{P}$. % orthogonal transformation
We recast \eqref{increment_1st_term} as
\begin{align} \label{increment_1}
	&\eta  \mbf{H} \nabla R \left(\mbf{V}^t\right) \mbf{\Sigma}^{\frac{1}{2}} 
	=2\eta \alpha \mbf{P} \mbf{\Lambda}^2 \mbf{E}^t \mbf{G}^t \mbf{\Sigma},
\end{align}
where $\mbf{E}^t = \left(\gamma\mbf{I} + \alpha \mbf{G}^t \left(\mbf{G}^t\right)^{\sf H} \right)^{-1}$.
By the Woodbury matrix identity \cite{horn2012matrix}, $\mbf{E}^t$ can be rewritten as
\begin{align} \label{orthogonal_projection}
	\mbf{E}^t = \frac{1}{\gamma} \left(\mbf{I} -\mbf{G}^t \left(\frac{\gamma}{\alpha} \mbf{I} + \left(\mbf{G}^t\right)^{\sf H} \mbf{G}^t\right)^{-1} \left(\mbf{G}^t\right)^{\sf H} \right).
\end{align}
From \eqref{orthogonal_projection}, 
%Notice that 
we observe that $\mbf{E}^t$ is approximately the orthogonal complement projector of $\mbf{G}^t$.
Consequently, the columns of $\mbf{E}^t \mbf{G}^t \mbf{\Sigma}$, roughly speaking, lie in the orthogonal complement of $\mathcal{R} \left(\mbf{G}^t\right)$.
From \eqref{increment_1}, the increment $\eta \mbf{H} \nabla R \left(\mbf{V}^t\right) \mbf{\Sigma}^{\frac{1}{2}}$ operates in the column space of $\mbf{P}$ to move the received noise-free features towards their orthogonal complement, which thus expands the volume of the received noise-free features so that the overall coding rate increases. %  (represented by $\mbf{G}^t$)
Please see Fig. \ref{GA} for an illustration.
We also note that $\mbf{\Lambda}^2$ in \eqref{increment_1} is merely a scaling factor of the orthonormal basis $\mbf{P}$, which does not change the overall expanding effect. % coordinates defined by the columns of $\mbf{P}$

The term $- \eta \mbf{H} \nabla R_c \left(\mbf{V}^t\right) \mbf{\Sigma}^{\frac{1}{2}}$ in \eqref{increment} corresponds to the increment brought by $\nabla R_c\left(\mbf{V}^t\right)$.
With \eqref{gradient_2}, we obtain
\begin{align} \label{increment_2}
	\!-\eta \mbf{H} \nabla R_c \left(\mbf{V}^t\right) \mbf{\Sigma}^{\frac{1}{2}}
	&= -2\eta \alpha \sum_{j \in \mathcal{J}} p_j \mbf{H} \mbf{H}^{\sf H} \mbf{D}_j^t \mbf{H} \mbf{V}^t \mbf{\Sigma}_j \mbf{\Sigma}^{\frac{1}{2}} \nonumber \\
	&=-2\eta \alpha \sum_{j \in \mathcal{J}} p_j\mbf{P} \mbf{\Lambda}^2 \mbf{C}^t_j \mbf{G}^t_j \mbf{\Sigma}_j^{\frac{1}{2}} \mbf{\Sigma}^{\frac{1}{2}},
\end{align}
where $\mbf{D}_j^t \triangleq \left(\gamma\mbf{I} + \alpha \mbf{H}\mbf{V}^t \mbf{\Sigma}_j \left(\mbf{V}^t\right)^{\sf H} \mbf{H}^{\sf H}\right)^{-1}$, and $\mbf{C}^t_j \triangleq \left(\gamma\mbf{I} + \alpha \mbf{G}^t_j \left(\mbf{G}^t_j\right)^{\sf H} \right)^{-1}$ with $\mbf{G}_j^t \triangleq \mbf{\Lambda} \mbf{Q}^{\sf H} \mbf{V}^t  \mbf{\Sigma}_j^{\frac{1}{2}}$ being the representation of $\mbf{H} \mbf{V}^t \mbf{\Sigma}_j^{\frac{1}{2}}$ under the orthonormal basis $\mbf{P}$.
Note that the columns of $\mbf{H} \mbf{V}^t \mbf{\Sigma}_j^{\frac{1}{2}}$ span the subspace of the received noise-free features of class $j$.
Similar to $\mbf{E}^t$, $\mbf{C}_j^t$ is approximately the orthogonal complement projector of $\mbf{G}^t_j$.
Thus, $\mbf{C}^t_j \mbf{G}^t_j \mbf{\Sigma}^{\frac{1}{2}}_j \mbf{\Sigma}^{\frac{1}{2}}$ represents the ``residuals'' of the received noise-free features of class $j$ that deviate from the subspace of class $j$.
This part of each class needs to be suppressed, hence the negative sign in \eqref{increment_2}.
To conclude, the increment $- \eta \mbf{H} \nabla R_c \left(\mbf{V}^t\right) \mbf{\Sigma}^{\frac{1}{2}}$ operates in the column space of $\mbf{P}$ to compress the received noise-free features of each class, so that the coding rate of each class reduces (see Fig. \ref{GA}).

%\color{blue}
\begin{remk}
	The derivation in this section shows that, following the gradient ascent direction of the precoder incrementally expands the overall volume spanned by different classes of the received features.
	In the meanwhile, it incrementally compresses the volume spanned by each individual class.
	The joint effect of expanding the whole and compressing each parts leads to a higher MCR$^2$ value.
\end{remk}
%\color{black}

%\newcounter{TempEqCnt} %??????TempEqCnt
%\setcounter{TempEqCnt}{\value{equation}} %?????????TempEqCnt
%\setcounter{equation}{36}
%\begin{figure*}[b]
%	\hrulefill
%	\begin{subequations}
%		\begin{align}
%			\text{(P2)}~\max_{\{\mbf{W}_j\}_{j\in \mathcal{J}_0}, \mbf{U}, \{\mbf{V}_k\}_{k\in \mathcal{K}}}  & \log \det \left(\mbf{W}_0\right) - \mathrm{tr} \left(\mbf{W}_0 \mbf{F}_0\left(\mbf{U}, \mbf{V}\right)\right) 
%			+ \sum_{j\in \mathcal{J}} p_j \big( \log \det \left(\mbf{W}_j\right) - \mathrm{tr} \left(\mbf{W}_j \mbf{F}_j\left(\mbf{V}\right)\right) \big)  \\ 
%			\operatorname{ s.t. } \quad ~~~~~~
%			& \frac{1}{T}\mathrm{tr} \left(\mbf{V}_k\mbf{\Sigma}^{(kk)} \mbf{V}_k^{\sf H}\right) \leq P_k, ~~k \in \mathcal{K}.
%		\end{align}
%	\end{subequations}
%\end{figure*}
%\setcounter{equation}{\value{TempEqCnt}}

\section{Algorithm Design for MCR$^2$ Based Precoding Optimization} \label{Section_Algorithm}
In this section, we propose an equivalent formulation of (P1) by introducing auxiliary variables. % , termed as MCR$^2$ transform,
Based on the equivalent formulation, we develop a BCA-type algorithm to efficiently solve the problem.
\subsection{Equivalent Formulation of (P1)}
\begin{lemm} \label{lemm}
%	Let $\mbf{W}_0\succ \mbf{0}$ and $\mbf{U}$ be the auxiliary variables.
	For any positive definite matrix $\mbf{F} \in \mathbb{C}^{r \times r}$,
	we have
	\begin{align} \label{lemm-1a}
		 -\log \det \left(\mbf{F} \right)
		 = \max_{\mbf{W}\succ \mbf{0}} ~\log \det \left(\mbf{W}\right) - \mathrm{tr} \left(\mbf{W} \mbf{F}\right) + r ,
	\end{align}
	where the positive definite matrix $\mbf{W} \in \mathbb{C}^{r \times r}$ is an auxiliary variable.
	The maximum of the right-hand side (r.h.s.) of \eqref{lemm-1a} is attained at
	\begin{align} \label{lemm-1b}
		\mbf{W}^\star = \mbf{F}^{-1}.
	\end{align}
\end{lemm}
\begin{IEEEproof}
	Since the problem in \eqref{lemm-1a} is an unconstrained convex problem, setting its first-order derivative with respect to (w.r.t.) $\mbf{W}$ to zero matrix yields the optimal $\mbf{W}$ in \eqref{lemm-1b}.
	The equality in \eqref{lemm-1a} can be verified by substituting \eqref{lemm-1b} into the r.h.s. of \eqref{lemm-1a}.
\end{IEEEproof}

\begin{lemm} \label{lemm2}
	Given matrix $\mbf{A} \in \mathbb{C}^{s \times l}$ and any positive definite matrix $\mbf{B} \in \mathbb{C}^{l \times l}$, we have
	\begin{align}\label{lemm-2}
		&\log \det \left(\mbf{I} + \mbf{A} \mbf{B} \mbf{A}^{\sf H}\right) \nonumber \\
		=\,& \max_{\mbf{\Omega} \succ \mbf{0}, \mbf{\Xi}} ~ \log \det \left(\mbf{\Omega}\right) - \mathrm{tr} \left(\mbf{\Omega} \mbf{K}\left(\mbf{\Xi}\right)\right) + l,
	\end{align}
	where $\mbf{\Xi} \in \mathbb{C}^{s \times l}$ and the positive definite matrix $\mbf{\Omega} \in \mathbb{C}^{l \times l}$ are two auxiliary variables, and
	\begin{align}
		\mbf{K} \left(\mbf{\Xi}\right) \triangleq \left(\mbf{I} - \mbf{\Xi}^{\sf H} \mbf{A} \mbf{B}^{\frac{1}{2}} \right) \left(\mbf{I} - \mbf{\Xi}^{\sf H} \mbf{A} \mbf{B}^{\frac{1}{2}}\right)^{\sf H} + \mbf{\Xi}^{\sf H} \mbf{\Xi}
	\end{align}
	is an $l$ by $l$ matrix function.
	The maximum of \eqref{lemm-2} is attained at
	\begin{align}
		\mbf{\Xi}^{\star} & = \left(\mbf{I}+\mbf{A}\mbf{B}\mbf{A}^{\sf H}\right)^{-1} \mbf{A} \mbf{B}^{\frac{1}{2}}, \\
		\mbf{\Omega}^{\star} & = \left(\mbf{K} \left(\mbf{\Xi}^\star\right)\right)^{-1} = \left(\mbf{I} - \left(\mbf{\Xi}^{\star}\right)^{\sf H} \mbf{A} \mbf{B}^{\frac{1}{2}}\right)^{-1}.
	\end{align}
%	\begin{align}
%		\mbf{\Xi}^{\star} = \left(\mbf{I}+\mbf{A}\mbf{B}\mbf{A}^{\sf H}\right)^{-1} \mbf{A} \mbf{B}^{\frac{1}{2}}
%	\end{align}
%	and
%	\begin{align}
%		\mbf{\Omega}^{\star} = \left(\mbf{K} \left(\mbf{\Xi}^\star\right)\right)^{-1} = \left(\mbf{I} - \left(\mbf{\Xi}^{\star}\right)^{\sf H} \mbf{A} \mbf{B}^{\frac{1}{2}}\right)^{-1}.
%	\end{align}
\end{lemm}
\begin{IEEEproof}
	The proof directly follows the principle behind the WMMSE framework \cite{shi2011iteratively}. 
	Here, we omit the details for brevity.
%	The details are omitted for brevity.	
\end{IEEEproof}

%By applying Lemmas \ref{lemm} and \ref{lemm2} to the second (summation) and first terms of the MCR$^2$ objective, respectively, we obtain the following theorem on the equivalent transform of (P1). % $R_c \left(\mbf{V}_1, \dots, \mbf{V}_K\right)$ and $R \left(\mbf{V}_1, \dots, \mbf{V}_K\right)$
%
%\begin{theo} \label{mcr2_transform}
%	(P1) is equivalent to (P2) (shown at the bottom of this page)
%	in the sense that the global optimal solution $\{\mbf{V}_k\}_{k\in \mathcal{K}}$ for the two problems are identical, where $\left\{\mbf{W}_j \succ \mbf{0}\right\}_{j\in \mathcal{J}_0}$ and $\mbf{U}$ are the auxiliary variables, $\mathcal{J}_0 \triangleq \left\{0\right\} \cup \mathcal{J}$;
%	$\mbf{F}_0 \left(\mbf{U}, \mbf{V}\right)$ and $\mbf{F}_j \left(\mbf{V}\right)$ are respectively given by
%	\setcounter{TempEqCnt}{\value{equation}} %?????????TempEqCnt
%	\setcounter{equation}{37}
%		\begin{align}
%			\mbf{F}_0 \left(\mbf{U}, \mbf{V}\right) &\triangleq \left(\mbf{I} - \mbf{U}^{\sf H} \mbf{H} \mbf{V} \mbf{\Sigma}^{\frac{1}{2}} \right) \left(\mbf{I} - \mbf{U}^{\sf H} \mbf{H} \mbf{V} \mbf{\Sigma}^{\frac{1}{2}}\right)^{\sf H} + \frac{\gamma}{\alpha} \mbf{U}^{\sf H} \mbf{U}, \label{E_0}\\
%			\mbf{F}_j \left(\mbf{V}\right) &\triangleq \gamma\mbf{I} + \alpha \mbf{H}\mbf{V} \mbf{\Sigma}_j \mbf{V}^{\sf H} \mbf{H}^{\sf H} , ~~ j\in \mathcal{J}. \label{E_j}
%		\end{align}
%\end{theo}

Define the optimization problem (P2) as
%By applying Lemmas 1 and 2 to $R_c \left(\mbf{V}_1, \dots, \mbf{V}_K\right)$ and $R \left(\mbf{V}_1, \dots, \mbf{V}_K\right)$ in \eqref{mcr2-precoding}, respectively, we obtain the following theorem on the equivalent transform of (P1). % $R_c \left(\mbf{V}_1, \dots, \mbf{V}_K\right)$ and $R \left(\mbf{V}_1, \dots, \mbf{V}_K\right)$
%By applying Lemmas 1 and 2 to $R_c \left(\mbf{V}_1, \dots, \mbf{V}_K\right)$ and $R \left(\mbf{V}_1, \dots, \mbf{V}_K\right)$ in \eqref{mcr2-precoding}, respectively, we obtain the following transform of (P1): % $R_c \left(\mbf{V}_1, \dots, \mbf{V}_K\right)$ and $R \left(\mbf{V}_1, \dots, \mbf{V}_K\right)$
%\begin{theo} \label{mcr2_transform}
%	(P1) is equivalent to
\begin{align}
	\text{(P2)}~\max_{\{\mbf{W}_j \succ \mbf{0}\}_{j\in \mathcal{J}_0}, \mbf{U}, \atop \{\mbf{V}_k \in \mathcal{V}_k\}_{k\in \mathcal{K}} }  & \log \det \left(\mbf{W}_0\right) - \mathrm{tr} \left(\mbf{W}_0 \mbf{F}_0\left(\mbf{U}, \mbf{V}\right)\right)  \nonumber \\
	&\!\!\!\! \!\!\!\! \!\!\!\! \!\! \!\!\!\! \!\! \!\!\! \!\! \! + \sum_{j\in \mathcal{J}} p_j \big( \log \det \left(\mbf{W}_j\right) - \mathrm{tr} \left(\mbf{W}_j \mbf{F}_j\left(\mbf{V}\right)\right) \big) ,
\end{align}
where $\left\{\mbf{W}_j \succ \mbf{0}\right\}_{j\in \mathcal{J}_0}$ and $\mbf{U}$ are the auxiliary variables, $\mathcal{J}_0 \triangleq \left\{0\right\} \cup \mathcal{J}$;
$\mbf{F}_0 \left(\mbf{U}, \mbf{V}\right)$ and $\mbf{F}_j \left(\mbf{V}\right)$ are respectively given by
%		\begin{align}
	%			\mbf{F}_0 \left(\mbf{U}, \mbf{V}\right) &\triangleq \left(\mbf{I} - \mbf{U}^{\sf H} \mbf{H} \mbf{V} \mbf{\Sigma}^{\frac{1}{2}} \right) \left(\mbf{I} - \mbf{U}^{\sf H} \mbf{H} \mbf{V} \mbf{\Sigma}^{\frac{1}{2}}\right)^{\sf H} + \frac{\gamma}{\alpha} \mbf{U}^{\sf H} \mbf{U}, \label{E_0}\\
	%			\mbf{F}_j \left(\mbf{V}\right) &\triangleq \gamma\mbf{I} + \alpha \mbf{H}\mbf{V} \mbf{\Sigma}_j \mbf{V}^{\sf H} \mbf{H}^{\sf H} , ~~ j\in \mathcal{J}. \label{E_j}
	%		\end{align}
\begin{align}
	\!\! \mbf{F}_0  &\triangleq \left(\mbf{I} - \mbf{U}^{\sf H} \mbf{H} \mbf{V} \mbf{\Sigma}^{\frac{1}{2}} \right) \! \left(\mbf{I} - \mbf{U}^{\sf H} \mbf{H} \mbf{V} \mbf{\Sigma}^{\frac{1}{2}}\right)^{\sf H} \! + \frac{\gamma}{\alpha} \mbf{U}^{\sf H} \mbf{U}, \label{E_0}\\
	\!\! \mbf{F}_j  &\triangleq \gamma\mbf{I} + \alpha \mbf{H}\mbf{V} \mbf{\Sigma}_j \mbf{V}^{\sf H} \mbf{H}^{\sf H} , ~~ j\in \mathcal{J}. \label{E_j}
\end{align}
%	\begin{align}
	%		\mbf{F}_0  &\triangleq \left(\mbf{I} - \mbf{U}^{\sf H} \mbf{H} \mbf{V} \mbf{\Sigma}^{\frac{1}{2}} \right)  \left(\mbf{I} - \mbf{U}^{\sf H} \mbf{H} \mbf{V} \mbf{\Sigma}^{\frac{1}{2}}\right)^{\sf H}  + \frac{\gamma}{\alpha} \mbf{U}^{\sf H} \mbf{U}, \nonumber\\
	%		\mbf{F}_j  &\triangleq \gamma\mbf{I} + \alpha \mbf{H}\mbf{V} \mbf{\Sigma}_j \mbf{V}^{\sf H} \mbf{H}^{\sf H} , ~~ j\in \mathcal{J}. \nonumber
	%	\end{align}
\begin{prop}
	(P1) is equivalent to (P2) in the sense that the optimal solution $\{\mbf{V}_k\}_{k\in \mathcal{K}}$ for the two problems are identical.
\end{prop}
\begin{IEEEproof}
	The equivalence can be verified by applying Lemmas 1 and 2 to $R_c \left(\mbf{V}_1, \dots, \mbf{V}_K\right)$ and $R \left(\mbf{V}_1, \dots, \mbf{V}_K\right)$ in \eqref{mcr2-precoding}, respectively.
\end{IEEEproof}

%\setcounter{TempEqCnt}{\value{equation}} %?????????TempEqCnt
%\setcounter{equation}{42}
%\begin{figure*}[b]
%	\hrulefill
%	\begin{subequations}
%		\begin{align}
%			\text{(P3)}~\min_{\left\{\mbf{V}_k\right\}_{k\in \mathcal{K}}} ~ &  
%			\mathrm{tr} \Bigg(\mbf{W}_0 \bigg(\mbf{I} - \mbf{U}^{\sf H} \sum_{p\in\mathcal{K}}\mbf{H}_p \mbf{V}_p \big(\mbf{\Sigma}^{\frac{1}{2}}\big)^{(p)} \bigg) \bigg(\mbf{I} - \mbf{U}^{\sf H} \sum_{q\in\mathcal{K}} \mbf{H}_q \mbf{V}_q \big( \mbf{\Sigma}^{\frac{1}{2}}\big)^{(q)}\bigg)^{\sf H} \Bigg) \nonumber \\
%			&+ \alpha \sum_{j\in \mathcal{J}} p_j \mathrm{tr} \Bigg(\mbf{W}_j \sum_{q\in\mathcal{K}} \bigg(\sum_{p\in\mathcal{K}} \mbf{H}_p \mbf{V}_p \mbf{\Sigma}_{j}^{(pq)}\bigg) \mbf{V}_q^{\sf H} \mbf{H}_q^{\sf H}\Bigg) \\ 
%			\operatorname{ s.t. } ~~~
%			& \frac{1}{T}\mathrm{tr} \left(\mbf{V}_k\mbf{\Sigma}^{(kk)} \mbf{V}_k^{\sf H}\right) \leq P_k, ~~k \in \mathcal{K}.
%		\end{align}
%	\end{subequations}
%\end{figure*}
%\setcounter{equation}{\value{TempEqCnt}}
\subsection{BCA Algorithm}
We propose a BCA algorithm to solve (P2), which updates one block of variables at a time with the other blocks fixed.
For any fixed $\left\{\mbf{W}_j\right\}_{j\in \mathcal{J}_0}$ and $\{\mbf{V}_k\}_{k\in \mathcal{K}}$, (P2) is an unconstrained convex optimization problem w.r.t. $\mbf{U}$.
%Therefore, by setting the first-order derivative w.r.t. $\mbf{U}$ to zero matrix, 
From Lemma \ref{lemm2},
we obtain the optimal solution
\begin{align} \label{opt-U}
	\mbf{U}^{\star} = \left(\mbf{H}\mbf{V}\mbf{\Sigma}\mbf{V}^{\sf H}\mbf{H}^{\sf H} +   \frac{\gamma}{\alpha} \mbf{I}\right)^{-1} \mbf{H} \mbf{V} \mbf{\Sigma}^{\frac{1}{2}} . 
\end{align}
For any fixed $\mbf{U}$ and $\{\mbf{V}_k\}_{k\in \mathcal{K}}$, the optimization w.r.t. each $\mbf{W}_j$ is also convex and can be decoupled across different $j$'s.
The optimal $\mbf{W}_j$ is given by
\begin{align} \label{opt-W}
	\mbf{W}_j^\star =  \mbf{F}_j^{-1}, ~~ j\in \mathcal{J}_0.
\end{align}

It remains to consider the update rule of $\left\{\mbf{V}_k\right\}_{k\in \mathcal{K}}$.
In particular, for any fixed $\mbf{U}$ and $\{\mbf{W}_j\}_{j\in \mathcal{J}_0}$, (P2) reduces to
\begin{align}
	\min_{\left\{\mbf{V}_k \in \mathcal{V}_k\right\}_{k\in \mathcal{K}}}~  &  \mathrm{tr} \left(\mbf{W}_0 \left(\mbf{I} - \mbf{U}^{\sf H} \mbf{H} \mbf{V} \mbf{\Sigma}^{\frac{1}{2}} \right) \left(\mbf{I} - \mbf{U}^{\sf H} \mbf{H} \mbf{V} \mbf{\Sigma}^{\frac{1}{2}}\right)^{\sf H} \right) \nonumber \\
	&+ \alpha \sum_{j\in \mathcal{J}} p_j \mathrm{tr} \left(\mbf{W}_j \mbf{H}\mbf{V} \mbf{\Sigma}_j \mbf{V}^{\sf H} \mbf{H}^{\sf H}\right).
\end{align}
%\begin{subequations}
%	\begin{align}
%		\min_{\left\{\mbf{V}_k\right\}_{k\in \mathcal{K}}}~  &  \mathrm{tr} \left(\mbf{W}_0 \left(\mbf{I} - \mbf{U}^{\sf H} \mbf{H} \mbf{V} \mbf{\Sigma}^{\frac{1}{2}} \right) \left(\mbf{I} - \mbf{U}^{\sf H} \mbf{H} \mbf{V} \mbf{\Sigma}^{\frac{1}{2}}\right)^{\sf H} \right) \nonumber \\
%		&+ \alpha \sum_{j\in \mathcal{J}} p_j \mathrm{tr} \left(\mbf{W}_j \mbf{H}\mbf{V} \mbf{\Sigma}_j \mbf{V}^{\sf H} \mbf{H}^{\sf H}\right) \\ 
%		\operatorname{ s.t. } ~~~
%		& \frac{1}{T}\mathrm{tr} \left(\mbf{V}_k\mbf{\Sigma}^{(kk)} \mbf{V}_k^{\sf H}\right) \leq P_k, ~~k \in \mathcal{K}.
%	\end{align}
%\end{subequations}
By noting that $\mbf{H} = \left[\mbf{H}_1 , \dots, \mbf{H}_K\right]$ and $\mbf{V} = \diag \left\{\mbf{V}_1, \dots, \mbf{V}_K\right\}$, we rewrite the above problem as
% (P3), shown at the bottom of this page.
\begin{align}
	\text{(P3)}~\min_{\left\{\mbf{V}_k \in \mathcal{V}_k\right\}_{k\in \mathcal{K}}} ~ &  
	\mathrm{tr} \bigg(\mbf{W}_0 \bigg(\mbf{I} - \mbf{U}^{\sf H} \sum_{p\in\mathcal{K}}\mbf{H}_p \mbf{V}_p \big(\mbf{\Sigma}^{\frac{1}{2}}\big)^{(p)} \bigg) \nonumber \\ 
	&\bigg(\mbf{I} - \mbf{U}^{\sf H} \sum_{q\in\mathcal{K}} \mbf{H}_q \mbf{V}_q \big( \mbf{\Sigma}^{\frac{1}{2}}\big)^{(q)}\bigg)^{\sf H} \bigg) \nonumber \\
	& \!\!\!\!\!\!\!\!\!\!\!\!\!\!\!\!\!\!\!\!\!\!\!\!\!\!\!\!\! \!\!\!\!\!\!\!\!\!\!\! \!\!\!\!\! + \alpha \sum_{j\in \mathcal{J}} p_j \mathrm{tr} \Bigg(\mbf{W}_j \sum_{q\in\mathcal{K}} \bigg(\sum_{p\in\mathcal{K}} \mbf{H}_p \mbf{V}_p \mbf{\Sigma}_{j}^{(pq)}\bigg) \mbf{V}_q^{\sf H} \mbf{H}_q^{\sf H}\Bigg). 
\end{align}
%\begin{align}
%	\min_{\left\{\mbf{V}_k \in \mathcal{V}_k\right\}_{k\in \mathcal{K}}}~  &  \mathrm{tr} \left(\mbf{W}_0 \left(\mbf{I} - \mbf{U}^{\sf H} \mbf{H} \mbf{V} \mbf{\Sigma}^{\frac{1}{2}} \right) \left(\mbf{I} - \mbf{U}^{\sf H} \mbf{H} \mbf{V} \mbf{\Sigma}^{\frac{1}{2}}\right)^{\sf H} \right) \nonumber \\
%	&+ \alpha \sum_{j\in \mathcal{J}} p_j \mathrm{tr} \left(\mbf{W}_j \mbf{H}\mbf{V} \mbf{\Sigma}_j \mbf{V}^{\sf H} \mbf{H}^{\sf H}\right) .
%\end{align}
%By noting that $\mbf{H} = \left[\mbf{H}_1 , \dots, \mbf{H}_K\right]$ and $\mbf{V} = \diag \left\{\mbf{V}_1, \dots, \mbf{V}_K\right\}$, we rewrite the above problem as (P3), shown at the bottom of this page.
In (P3), $\big(\mbf{\Sigma}^{\frac{1}{2}}\big)^{(p)}$ denotes row $m_p$ to row $n_p$ of $\mbf{\Sigma}^{\frac{1}{2}}$, where $m_p \triangleq \sum_{i=0}^{p-1}\frac{D_i}{2} +1$ and $n_p \triangleq \sum_{i=1}^p \frac{D_i}{2}$, with $D_0 \triangleq 0$;
%$\mbf{\Sigma}_{j}^{(pq)} \triangleq \mathbb{E} \left[\dot{\mbf{z}}_p \dot{\mbf{z}}_q^{\sf H}\right]-\bsm{\mu}_p \bsm{\mu}_q^{\sf H}$ is the covariance matrix between $\dot{\mbf{z}}_p$ and $\dot{\mbf{z}}_q$.
$\mbf{\Sigma}_{j}^{(pq)}$ is formed by row $m_p$ to row $n_p$ and column $m_q$ to column $n_q$ of $\mbf{\Sigma}_j$. %, where $u \triangleq \sum_{i=0}^{q-1}D_i/2 +1$ and $v \triangleq \sum_{i=1}^q D_i/2$.
It is seen that the precoder $\mbf{V}_k$ at device $k$ is coupled with the other precoders in the objective function of (P3).
Nevertheless, alternatingly optimizing one of the precoder with the others fixed results in a convex quadratically constrained quadratic program (QCQP). %, which can be solved by standard convex optimization algorithms, e.g., interior-point method \cite{boyd2004convex}.

%In (P3), $\big(\mbf{\Sigma}^{\frac{1}{2}}\big)^{(p)}$ denotes row $m$ to row $n$ of $\mbf{\Sigma}^{\frac{1}{2}}$, where $m \triangleq \sum_{i=0}^{p-1}D_i/2 +1$ and $n \triangleq \sum_{i=1}^p D_i/2$, with $D_0 \triangleq 0$;
%$\mbf{\Sigma}_{j}^{(pq)} \triangleq \mathbb{E} \left[\dot{\mbf{z}}_p \dot{\mbf{z}}_q^{\sf H}\right]-\bsm{\mu}_p \bsm{\mu}_q^{\sf H}$ is the matrix formed by row $m$ to row $n$ and column $u$ to column $v$ of $\mbf{\Sigma}_j$, where $u \triangleq \sum_{i=0}^{q-1}D_i/2 +1$ and $v \triangleq \sum_{i=1}^q D_i/2$.
%It is seen that the precoder $\mbf{V}_k$ at device $k$ is coupled with the other precoders in the objective function of (P3).
%Nevertheless, alternatingly optimizing one of the precoder with the others fixed results in a convex quadratically constrained quadratic program (QCQP). %, which can be solved by standard convex optimization algorithms, e.g., interior-point method \cite{boyd2004convex}.
%To show it more explicitly, we write the sub-problem for updating $\mbf{V}_k$ in (P4), shown at the bottom of this page.
%In (P4), the terms irrelevant to the optimization of $\mbf{V}_k$ are omitted.

To see this more clearly, we write the sub-problem for updating $\mbf{V}_k$ in (P4) by ignoring the terms irrelevant to optimization as
\begin{subequations}
	\begin{align}
		\text{(P4)}~~\min_{\mbf{V}_k} ~~ &  
		-2 \mathrm{Re} \left\{ \mathrm{tr} \left( \mbf{T}_k^{\sf H} \mbf{V}_k \right) \right\} 
		+ \mathrm{tr} \left(\mbf{\Phi}_k \mbf{V}_k \mbf{\Sigma}^{(kk)} \mbf{V}_k^{\sf H} \right) \nonumber \\
		&+ \alpha \sum_{j\in\mathcal{J}} p_j \mathrm{tr} \left( \mbf{\Psi}_{j,k} \mbf{V}_k \mbf{\Sigma}_j^{(kk)} \mbf{V}_k^{\sf H}  \right)\\ 
		\operatorname{ s.t. } ~~
		& \mathrm{tr} \left(\mbf{V}_k\mbf{\Sigma}^{(kk)} \mbf{V}_k^{\sf H}\right) \leq T P_k,
	\end{align}
\end{subequations}
where $\mbf{T}_k \triangleq \mbf{T}_{k,1} - \mbf{T}_{k,2} - \mbf{T}_{k,3}$ with the three components defined as $\mbf{T}_{k,1} \triangleq \mbf{H}^{\sf H}_k \mbf{U} \mbf{W}_0 \left(\big(\mbf{\Sigma}^{\frac{1}{2}}\big)^{(k)} \right)^{\sf H}$, $\mbf{T}_{k,2} \triangleq \mbf{H}^{\sf H}_k \mbf{U} \mbf{W}_0 \mbf{U}^{\sf H}\sum_{q \neq k} \mbf{H}_q \mbf{V}_q \mbf{\Sigma}^{(qk)}$, and $\mbf{T}_{k,3} \triangleq \alpha \sum_{j\in\mathcal{J}} p_j \mbf{H}^{\sf H}_k \mbf{W}_j \sum_{q \neq k} \mbf{H}_q \mbf{V}_q \mbf{\Sigma}_j^{(qk)}$;
$\mbf{\Phi}_k  \triangleq \mbf{H}_k^{\sf H} \mbf{U} \mbf{W}_0 \mbf{U}^{\sf H} \mbf{H}_k$;
$\mbf{\Psi}_{j,k} \triangleq \mbf{H}_k^{\sf H}\mbf{W}_j \mbf{H}_k$.
%In (P4), the terms irrelevant to the optimization of $\mbf{V}_k$ are omitted.
Indeed, (P4) can be solved in closed form by exploiting its first-order optimality conditions.
For tractability, define $\mbf{v}_k \triangleq \mathrm{vec} \left(\mbf{V}_k\right)$ and $\mbf{t}_k \triangleq \mathrm{vec} \left(\mbf{T}_k\right)$.
By applying the facts $\mathrm{tr}\left(\mbf{A}^{\sf H} \mbf{B}\right) = \mathrm{vec}^{\sf H}\left(\mbf{A}\right) \mathrm{vec}\left(\mbf{B}\right)$ and $\mathrm{vec} \left(\mbf{A}\mbf{B}\right) = \left(\mbf{B}^{\sf T} \otimes \mbf{I}\right)\mathrm{vec} \left(\mbf{A}\right)$, we rewrite (P4) in a vectorized form as
%Indeed, (P4) can be solved in closed form by exploiting its first-order optimality conditions.
%Define $\mbf{v}_k \triangleq \mathrm{vec} \left(\mbf{V}_k\right)$ and $\mbf{t}_k \triangleq \mathrm{vec} \left(\mbf{T}_k\right)$.
%By applying the facts $\mathrm{tr}\left(\mbf{A}^{\sf H} \mbf{B}\right) = \mathrm{vec}^{\sf H}\left(\mbf{A}\right) \mathrm{vec}\left(\mbf{B}\right)$ and $\mathrm{vec} \left(\mbf{A}\mbf{B}\right) = \left(\mbf{B}^{\sf T} \otimes \mbf{I}\right)\mathrm{vec} \left(\mbf{A}\right)$, we rewrite (P4) in a vectorized form as
\begin{subequations}
	\begin{align}
		\text{(P5)}~~\min_{\mbf{v}_k} ~~&  
		-2 \mathrm{Re} \left\{ \mbf{t}_k^{\sf H} \mbf{v}_k \right\}
		+ \mbf{v}_k^{\sf H}  \mbf{M}_k \mbf{v}_k \\ 
		\operatorname{ s.t. } ~~
		& \mbf{v}_k^{\sf H} \left(\big(\mbf{\Sigma}^{(kk)}\big)^{\sf T}\otimes \mbf{I} \right) \mbf{v}_k\leq T P_k,
	\end{align}
\end{subequations}
%where $\mbf{T}_k$ and $\mbf{M}_k$ are respectively given in \eqref{T_k} and \eqref{M_k} at the bottom of the next page.
where $\mbf{M}_k \triangleq  \big(\mbf{\Sigma}^{(kk)}\big)^{\sf T} \otimes \mbf{\Phi}_k 
+ \alpha \sum_{j\in\mathcal{J}} p_j \big(\mbf{\Sigma}_j^{(kk)}\big)^{\sf T} \otimes \mbf{\Psi}_{j,k}$.
%\begin{align}
%	\mbf{T}_k \triangleq \;& \mbf{H}^{\sf H}_k \mbf{U} \mbf{W}_0 \left(\big(\mbf{\Sigma}^{\frac{1}{2}}\big)^{(k)} \right)^{\sf H}
%	- \mbf{H}^{\sf H}_k \mbf{U} \mbf{W}_0 \mbf{U}^{\sf H}\sum_{q \neq k} \mbf{H}_q \mbf{V}_q \mbf{\Sigma}^{(qk)} \nonumber \\
%	&- \alpha \sum_{j\in\mathcal{J}} p_j \mbf{H}^{\sf H}_k \mbf{W}_j \sum_{q \neq k} \mbf{H}_q \mbf{V}_q \mbf{\Sigma}_j^{(qk)}, \label{T_k}\\
%	\mbf{M}_k \triangleq \;&  \big(\mbf{\Sigma}^{(kk)}\big)^{\sf T} \otimes \big(\mbf{H}_k^{\sf H} \mbf{U} \mbf{W}_0 \mbf{U}^{\sf H} \mbf{H}_k \big) \nonumber \\
%	&+ \alpha \sum_{j\in\mathcal{J}} p_j \big(\mbf{\Sigma}_j^{(kk)}\big)^{\sf T} \otimes \big( \mbf{H}_k^{\sf H}\mbf{W}_j \mbf{H}_k \big). \label{M_k}
%\end{align}
By letting $\widetilde{\mbf{M}}_k \triangleq \big((\mbf{\Sigma}^{(kk)})^{\sf T}\otimes \mbf{I} \big)^{-\frac{1}{2}} \mbf{M}_k \big((\mbf{\Sigma}^{(kk)})^{\sf T}\otimes \mbf{I} \big)^{-\frac{1}{2}}$, 
$\widetilde{\mbf{t}}_k \triangleq \big((\mbf{\Sigma}^{(kk)})^{\sf T}\otimes \mbf{I} \big)^{-\frac{1}{2}} \mbf{t}_k$, and $\widetilde{\mbf{v}}_k \triangleq \big((\mbf{\Sigma}^{(kk)})^{\sf T}\otimes \mbf{I} \big)^{\frac{1}{2}} \mbf{v}_k$,
(P5) can be rewritten as
\begin{subequations}
	\begin{align}
		\text{(P6)}~~\min_{\widetilde{\mathbf{v}}_k} ~~ & -2  \mathrm{Re} \big\{\widetilde{\mbf{t}}_k^{\sf H} \widetilde{\mbf{v}}_k \big\}
		+
		\widetilde{\mbf{v}}_k^{\sf H} \widetilde{\mbf{M}}_k \widetilde{\mbf{v}}_k \\ 
		\operatorname{ s.t. } ~~
		& \widetilde{\mathbf{v}}_k^{\sf H} \widetilde{\mathbf{v}}_k \leq T P_k. \label{lagrangian-multiplier}
	\end{align}
\end{subequations}
%The Lagrangian function of (P6) is expressed as
%\begin{align}
%	L\left(\mbf{r}_k, \lambda_k\right) = -2  \mathrm{Re} \left\{\mbf{t}^{\sf H}_k \mbf{r}_k \right\} + \mbf{r}_k^{\sf H} \mbf{N}_k \mbf{r}_k
%	+ \lambda_k \left(\mbf{r}^{\sf H}_k \mbf{r}_k- P_kT\right),
%\end{align}
%where $\lambda_k$ denotes the dual variable associated with the constraint \eqref{lagrangian-multiplier}.
By the first-order optimality condition, the optimal solution to (P6) is given by %and rearranging the terms
\begin{align}
	\widetilde{\mbf{v}}_k^\star = \big(\widetilde{\mbf{M}}_k + \lambda_k \mbf{I}\big)^{-1} \widetilde{\mbf{t}}_k. \label{v-update}
\end{align}
In \eqref{v-update}, $\lambda_k$ denotes the dual variable associated with \eqref{lagrangian-multiplier}, which can be efficiently calculated via a bisection search.

We summarize the implementation details of the BCA algorithm in Algorithm \ref{alg1}.

\begin{algorithm}[t]\label{alg1}
	\SetAlgoLined
	%\SetKwInOut{Input}{Input}\SetKwInOut{Output}{Output}
	%	\SetKw{KwInput}{Input:}
	%	\SetKw{KwOutput}{Output:}
	%	\KwInput{
		%		$\left\{m_s^{\sf r}\right\}_{s=1}^{\min \{L_1, L_2\}}$, $\left\{m_i^{\sf d}\right\}_{i=1}^{L_3}$, $P$, grid size $s_{\sf grid}$, accuracy tolerance $\epsilon_{\sf acc}$;\
		%	}
	%	
	%	\KwOutput{
		%		Optimal solution pair $(\mbf{p}, \mbf{t})$\;
		%	}
	
	%	Initialize $\left\{\mbf{V}_k\right\}_{k\in \mathcal{K}}$ such that $\mathrm{tr} \left(\mbf{V}_k\mbf{\Sigma}^{(kk)} \mbf{V}_k^{\sf H}\right) = TP_k$;
	
	Initialize $\left\{\mbf{V}_k \in \mathcal{V}_k\right\}_{k\in \mathcal{K}}$;
	
	%	\For{$\mathcal{S}_{\sf a} \in \left\{\{1,2\},\cdots,\{1,2,\cdots, \color{blue} \min \{L_1, L_2\} \color{black} \} \right\}$}
	\While{convergence criterion is not met}
	{
		%		$\mbf{U} \leftarrow \left(\mbf{H}\mbf{V}\mbf{\Sigma}\mbf{V}^{\sf H}\mbf{H}^{\sf H} +   \frac{\gamma}{\alpha} \mbf{I}\right)^{-1} \mbf{H} \mbf{V} \mbf{\Sigma}^{\frac{1}{2}}$; 
		Update $\mbf{U}$ by \eqref{opt-U}; \label{U-Alg} \
		%		$\mbf{U} = \left(\mbf{H}\mbf{V}\mbf{\Sigma}\mbf{V}^{\sf H}\mbf{H}^{\sf H} +   \frac{\gamma}{\alpha} \mbf{I}\right)^{-1} \mbf{H} \mbf{V} \mbf{\Sigma}^{\frac{1}{2}}$; \
		
		%		$\mbf{W}_j \leftarrow \mbf{F}_j^{-1}$, $\forall j \in \mathcal{J}_0$; 
		Update $\mbf{W}_j$ by \eqref{opt-W}, $\forall j \in \mathcal{J}_0$; \label{W-Alg}\
		% , where $\mbf{F}_j$ is given in \eqref{E_0} and \eqref{E_j}
		
		\For{$k \in \mathcal{K}$}
		{
			%		$\mbf{r}_k \leftarrow \left(\mbf{N}_k + \lambda_k \mbf{I}\right)^{-1} \mbf{t}_k$;
			Update $\widetilde{\mbf{v}}_k$ by \eqref{v-update}; \label{r-Alg} \
			
			Retrive $\mbf{v}_k$ by $\mbf{v}_k = \big((\mbf{\Sigma}^{(kk)})^{\sf T}\otimes \mbf{I} \big)^{-\frac{1}{2}} \widetilde{\mbf{v}}_k$; \label{v-Alg}\
			
			Reshape $\mbf{v}_k$ in matrix form as $\mbf{V}_k$; 
		}
	}
	\caption{BCA Algorithm}
\end{algorithm}

%\begin{algorithm}[t]
%	\SetAlgoLined	
%	Initialize $\left\{\mbf{V}_k \in \mathcal{V}_k\right\}_{k\in \mathcal{K}}$;
%	
%	%	\For{$\mathcal{S}_{\sf a} \in \left\{\{1,2\},\cdots,\{1,2,\cdots, \color{blue} \min \{L_1, L_2\} \color{black} \} \right\}$}
%	\While{convergence criterion is not met}
%	{
%		$\mbf{U} \leftarrow \left(\mbf{H}\mbf{V}\mbf{\Sigma}\mbf{V}^{\sf H}\mbf{H}^{\sf H} +   \frac{\gamma}{\alpha} \mbf{I}\right)^{-1} \mbf{H} \mbf{V} \mbf{\Sigma}^{\frac{1}{2}} $; \
%		
%				$\mbf{W}_j \leftarrow \mbf{F}_j^{-1}$, $\forall j \in \mathcal{J}_0$; \
%		% , where $\mbf{F}_j$ is given in \eqref{E_0} and \eqref{E_j}
%		
%		\For{$k \in \mathcal{K}$}
%		{
%			$\widetilde{\mbf{v}}_k \leftarrow \big(\widetilde{\mbf{M}}_k + \lambda_k \mbf{I}\big)^{-1} \widetilde{\mbf{t}}_k$; \
%			
%			$\mbf{v}_k \leftarrow \big((\mbf{\Sigma}^{(kk)})^{\sf T}\otimes \mbf{I} \big)^{-\frac{1}{2}} \widetilde{\mbf{v}}_k$; \
%			
%			Reshape $\mbf{v}_k$ in matrix form as $\mbf{V}_k$; 
%		}
%	}
%	\caption{BCA Algorithm}
%\end{algorithm}

\subsection{Computational Complexity}
The computational complexity of the BCA algorithm is analyzed as follows.
In Line \ref{U-Alg} of Algorithm \ref{alg1}, the most computationally intensive operation lies in the matrix inversion, which has a complexity of $\mathcal{O}\left(T^3 N_{\sf r}^3\right)$. 
The operations in Line \ref{W-Alg} have a complexity of $\mathcal{O}\left(D^3 + J T^3 N_{\sf r}^3\right)$.
We assume that each device has the same number $N_{{\sf t}, k}$ of transmit antennas, and the same dimension $D_k$ of encoded features.
In Line \ref{r-Alg}, the complexity of calculating $\widetilde{\mbf{v}}_k$ is $\mathcal{O}\left(T^3 N_{{\sf t}, k}^3 D_k^3\right)$, in which the complexity of evaluating $\lambda_k$ is ignored.
In Line \ref{v-Alg}, calculating $\big((\mbf{\Sigma}^{(kk)})^{\sf T}\otimes \mbf{I} \big)^{-\frac{1}{2}}$ requires the SVD of $(\mbf{\Sigma}^{(kk)})^{\sf T}$ with a complexity of $\mathcal{O}\left(D_k^3\right)$, which, however, can be calculated offline and the complexity is therefore ignored. % prior to the execution of the algorithm
To summarize, the complexity of Algorithm \ref{alg1} is given by $\mathcal{O} \left(I_0 \left(D^3 + J T^3 N_{\sf r}^3 + I_1 K T^3 N_{{\sf t}, k}^3 D_k^3\right)\right)$, where $I_0$ and $I_1$ denote the numbers of outer and inner iterations, respectively.

\begin{figure}
	[t]
	\centering
	%	\vspace{-1em}
	\includegraphics[width=.85\columnwidth]{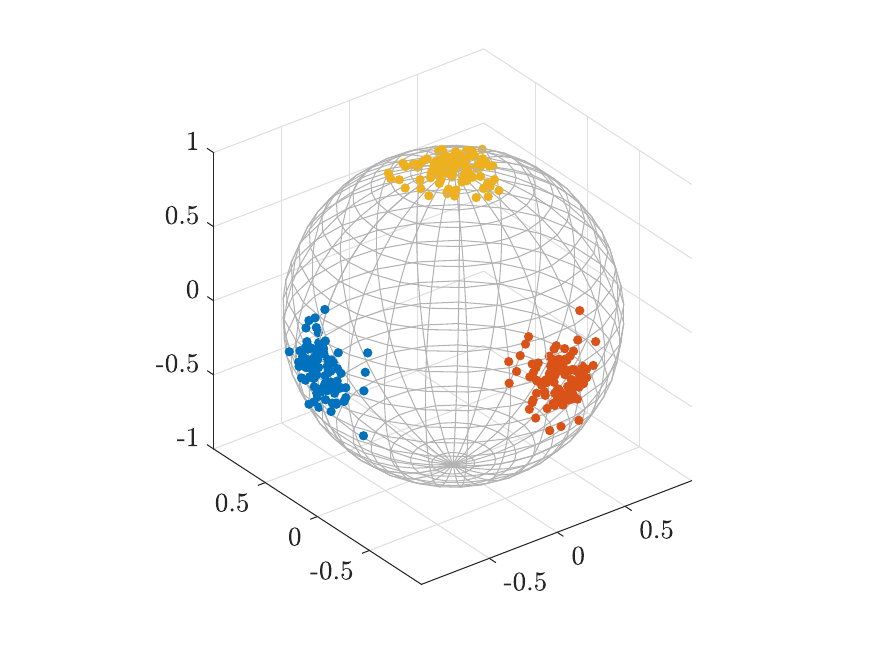}
	\vspace{-.5em}
	\caption{Synthetic feature samples generated from a GM distribution.}
	\label{sphere_sim}
	%	\vspace{-1em}
\end{figure}

\section{Simulation Results} \label{Section_Simulation}
%In this section, we present the experimental results of the MCR$^2$ based task-oriented communication design on both the synthetic features and real-world datasets.
\subsection{Results on Synthetic Features}
%\subsubsection{Dataset}
%In this subsection, we apply the MCR$^2$ precoder for feature transmission and visualize the received features using a synthetic feature dataset.
In this subsection, we visualize the received features obtained with MCR$^2$ precoding using a synthetic feature dataset.
The purpose of this subsection is not to compete with any other benchmarks, but to provide a deeper understanding on the mechanism of MCR$^2$ precoding.
Comparisons with various benchmarks on more complicated datasets are deferred to Section \ref{Simulation_II}.
The feature dataset is constructed by a mixture of three Gaussian distributions in $\mathbb{R}^3$ that is projected onto $\mathbb{S}^2$, as shown in Fig. \ref{sphere_sim}.
The feature samples from the three Gaussian components are labeled as three different classes, which are marked in orange, blue, and yellow.
In particular, we first generate a total of $300$ feature samples from the GM distribution $p\left(\mbf{z}\right) =\frac{1}{3} \sum_{j =1}^3 \mathcal{N} \left(\mbf{z} ; \bsm{\mu}_j, \mbf{\Sigma}_j\right)$,
where $\bsm{\mu}_1 = \left[-1, 0, 0\right]^{\sf T}$, $\bsm{\mu}_2 = \left[0, -1, 0\right]^{\sf T}$, $\bsm{\mu}_3 = \left[0, 0, 1\right]^{\sf T}$, and $\mbf{\Sigma}_1 = \mbf{\Sigma}_2 = \mbf{\Sigma}_3 = 0.02 \mbf{I}$.
Then, we project each feature sample onto the sphere $\mathbb{S}^2$ by $\mbf{z} / \left\|\mbf{z}\right\|_2$.

%\subsubsection{Communication Settings}
For better illustration, we consider a single-device case and assume a real-valued channel $\mbf{H} \in \mathbb{R}^{3 \times 3}$ for feature transmission.
Each entry of $\mbf{H}$ is modeled as i.i.d. Gaussian with zero mean and unit variance.
The precoder $\mbf{V} \in \mathbb{R}^{3 \times 3}$ and the AWGN $\mbf{n} \in \mathbb{R}^3$ are also real-valued. % assumed to be
The received feature $\mbf{y} \in \mathbb{R}^3$ is given by $\mbf{y} = \mbf{H} \mbf{V} \mbf{z} + \mbf{n}$.
We define the transmit SNR by $\mathrm{SNR}_{\sf t} = P_0 / \delta_0^2$, where $P_0$ denotes the maximum transmit power at the device.
%
%average received SNR as
%\begin{align}
%	\text{SNR} = \frac{P_0 \mathbb{E}\big[\left\|\mbf{H}\right\|_F^2\big]}{3\delta_0^2},
%\end{align}
%where $P_0$ denotes the maximum transmit power at the device.

\begin{figure*}
	[t]
	%	\vspace{-1em}
	\centering
	\subfigure[MCR$^2$ precoding]
	{\includegraphics[width=.95\columnwidth]{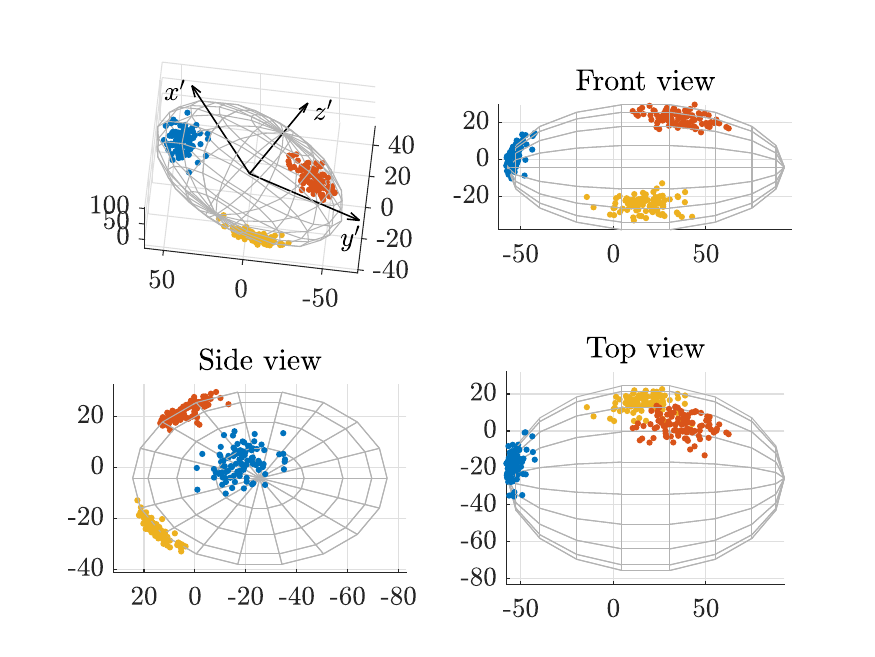}\label{ellipsoid_hp_c}}
	\subfigure[w/o precoding]
	{\includegraphics[width=.95\columnwidth]{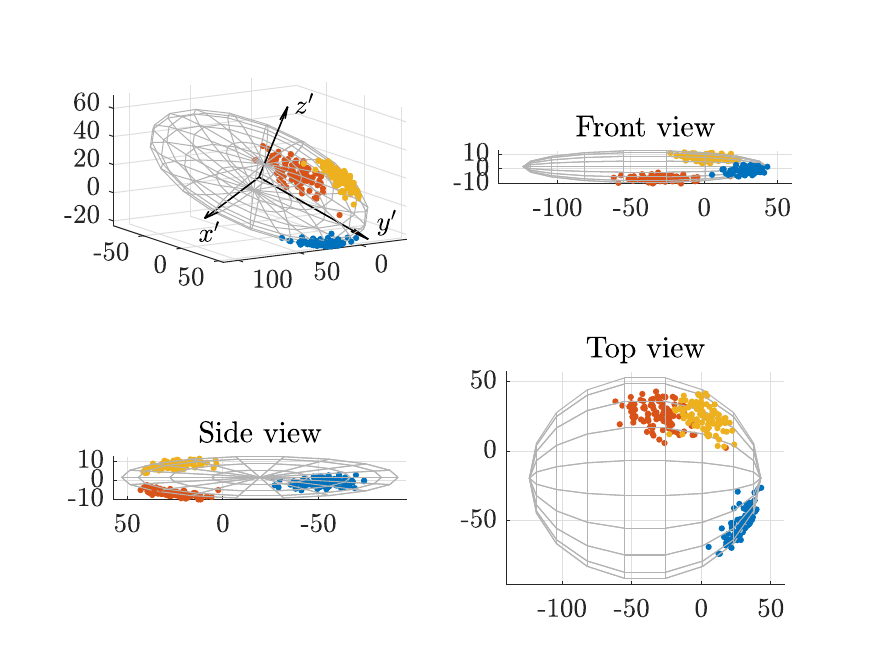}\label{ellipsoid_hp_d}}
	\caption{Visualization of the received feature samples on the synthetic feature dataset at $\mathrm{SNR}_{\sf t} = 30$ dB.
	The volume of the ellipsoid in (a) is $5.5350 \times 10^5$, larger than $2.8316 \times 10^5$ for that in (b).}
	\label{ellipsoid_hp}
	%	\vspace{-2em}
\end{figure*}
\begin{figure*}
	[t]
	%	\vspace{-1em}
	\centering
	\subfigure[MCR$^2$ precoding]
	{\includegraphics[width=.95\columnwidth]{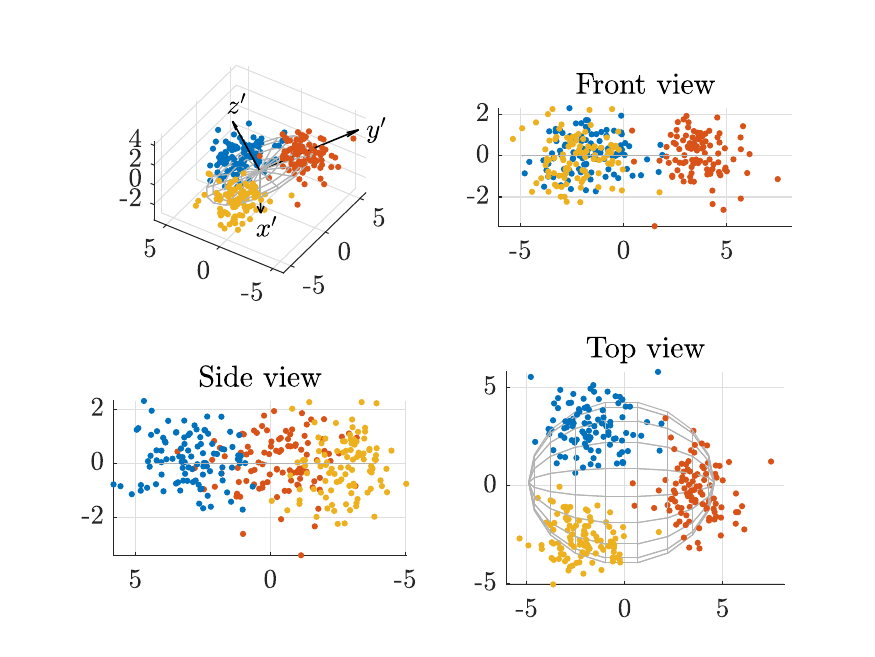}\label{ellipsoid_lp_c}}
	\subfigure[w/o precoding]
	{\includegraphics[width=.95\columnwidth]{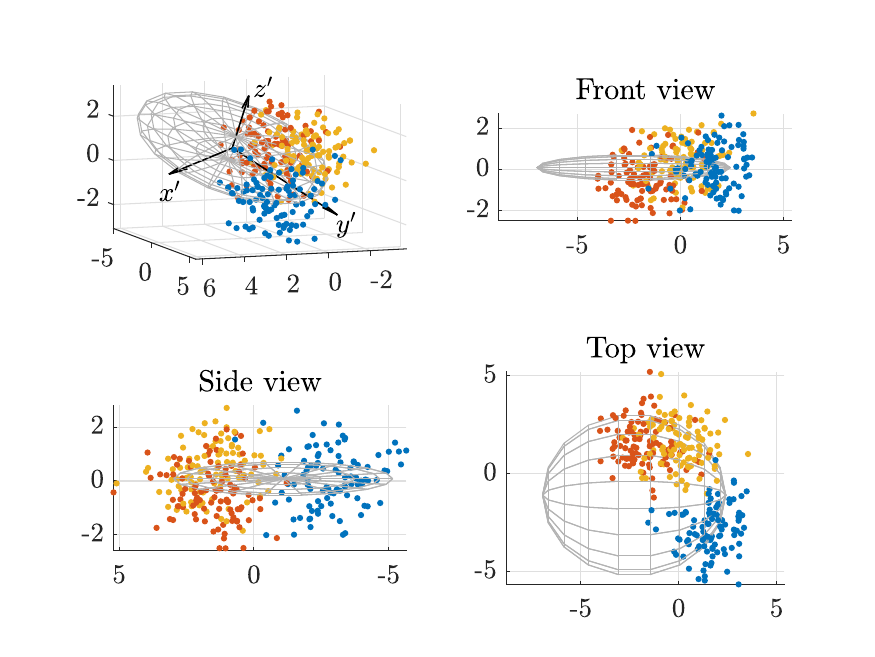}\label{ellipsoid_lp_d}}
	%	\caption{Lower Power regime: the volume of (a) is $2.2734$, and the volume of (b) is $6.2641 \times 10^{-14}$.}
	\caption{Visualization of the received feature samples on the synthetic feature dataset at $\mathrm{SNR}_{\sf t} = 5$ dB.
	The volume of the ellipsoid in (a) is $6.7130 \times 10^{-13}$, significantly smaller than $50.3539$ for that in (b).}
	\label{ellipsoid_lp}
	%	\vspace{-2em}
\end{figure*}

%In Fig. \ref{ellipsoid_hp}, we plot the received feature samples for a particular channel realization at $\mathrm{SNR}_{\sf t} = 30$ dB by applying the MCR$^2$ precoder and the random precoder.
In Fig. \ref{ellipsoid_hp_c}, we plot the received feature samples for a particular channel realization at $\mathrm{SNR}_{\sf t} = 30$ dB with MCR$^2$ precoding.
The results obtained without precoding is shown in Fig. \ref{ellipsoid_hp_d} by applying $\mbf{V} = \sqrt{P_0} \mbf{I}$. \color{black}
%The random precoder sets each element in $\mbf{V}$ randomly and then normalizes $\mbf{V}$ to satisfy the transmit power constraint.
In both subfigures, the ellipsoid in gray line represents the surface on which the received noise-free features are supposed to lie, with the ellipsoid equation given by $\mbf{y}^{\sf T} \left( \mbf{H}\mbf{V}\mbf{V}^{\sf T}\mbf{H}^{\sf T} \right)^{-1} \mbf{y} = 1$.
For better illustration, we define a new coordinate system $x^\prime$-$y^\prime$-$z^\prime$ whose axes are aligned with the principal axes of the ellipsoid.
We provide the front view, side view, and top view based on the $x^\prime$-$y^\prime$-$z^\prime$ coordinate system.
We observe that the different classes of received features are separated for both the MCR$^2$ precoding and without precoding schemes.
Notably, the ellipsoid in Fig. \ref{ellipsoid_hp_c} has a larger volume compared to that in Fig. \ref{ellipsoid_hp_d}.
In other words, the MCR$^2$ precoder usually leads to a larger distance between any two different received classes, resulting in a more robust performance against channel noise.

Fig. \ref{ellipsoid_lp} plots the received feature samples for the same channel realization at $\mathrm{SNR}_{\sf t} = 5$ dB. %a different SNR, i.e., $30$ dB.
Fig. \ref{ellipsoid_lp_c} shows that the ellipsoid is compressed to an ellipse,
%indicating that the received noise-free features obtained by applying the MCR$^2$ precoder roughly lie on a 2D surface.
indicating that the received noise-free features with MCR$^2$ precoding roughly lie on a 2D surface.
Nevertheless, the different classes of received features can still be roughly distinguished from its top view.
%In contrast, for the random precoder, different classes of the received features are overlapped in all the three views.
In contrast, the orange and yellow classes are overlapped in Fig. \ref{ellipsoid_lp_d} without precoding.
By comparing Fig. \ref{ellipsoid_hp_c} and Fig. \ref{ellipsoid_lp_c}, we observe that the MCR$^2$ precoder tends to expand the volume of the ellipsoid at high SNR, whereas tends to compress the volume at low SNR.
%spanned by the noise-free received features
The opposite behaviors at high and low SNRs are explained as follows.
Preserving a 3D ellipsoid requires a full-rank precoder which allocates the power to all the three eigen directions of the MIMO channel. 
At high SNR, owing to the abundant power resource, it is natural to pursue a full-rank precoder which expands the volume of the ellipsoid to separate different classes.
At low SNR, however, the limited power resource cannot afford to expand the volume large enough. %  for the purpose of classification
%Instead, the MCR$^2$ precoder tends to allocate power to only the two eigen directions with larger eigenvalues to expand the area of the spanned 2D surface, whereas the volume of the ellipsoid is compressed.
%%In this way, the three different classes can be correctly classified from a 2D view, shown in the top view in Fig. \ref{ellipsoid_lp_c}.
%In this way, the three different classes can still be correctly classified from a 2D view, shown in the top view in Fig. \ref{ellipsoid_lp_c}.
Instead, the MCR$^2$ precoder tends to allocate power to only the two eigen directions with larger eigenvalues to expand the area of the spanned 2D surface, such that the three different classes can be correctly classified from a 2D view.
Due to the absence of power on the channel's smallest eigenmode, the volume of the ellipsoid is hence compressed.
%In this way, the three different classes can be correctly classified from a 2D view, shown in the top view in Fig. \ref{ellipsoid_lp_c}.
%In this way, the three different classes can still be correctly classified from a 2D view, shown in the top view in Fig. \ref{ellipsoid_lp_c}.

\subsection{Results on CIFAR-10 and ModelNet10} \label{Simulation_II}
In this subsection, we consider the classification task on the CIFAR-10 \cite{krizhevsky2009learning} and ModelNet10 \cite{wu20153d} datasets for multi-device edge inference.
The CIFAR-10 dataset contains $60,000$ color images in 10 different classes, with $6,000$ images per class.
In particular, the training set and the testing set contain $5,000$ and $1,000$ images per class, respectively.
We assume that there are two devices, i.e., $K=2$, for the experiments on the CIFAR-10 dataset.
To generate two distinct views as the input to each device, following \cite{whang2021neural}, we split each image in the CIFAR-10 dataset vertically into two images of the same size.
For the experiments on the CIFAR-10 dataset, we adopt the ResNet18 \cite{he2016deep} as the backbone of the feature encoder at each device.
The ModelNet10 dataset comprises $10$ classes of computer-aided design (CAD) objects (e.g., sofa, bathtub, bed) with twelve views per sample.
The number of samples per class is imbalanced, ranging from $106$ to $889$ for training samples, and $50$ to $100$ for testing samples.
We assume that there are three devices, i.e., $K=3$, for the experiments on the ModelNet10 dataset.
The input views of the three devices are selected among the twelve views in the ModelNet10 dataset.
We adopt the VGG11 \cite{simonyan2014very} as the backbone of the feature encoder at each device.

%\subsubsection{System and Communication Settings}
The feature dimension at each device is set to $D_k = 8$ for the experiments on both the CIFAR-10 and ModelNet10 datasets.
Each edge device is equipped with $N_{{\sf t}, k} = 4$ transmit antennas.
The number of receive antennas is $N_{\sf r} = 8$ unless specified otherwise.
%The number of receive antennas is $N_{\sf r} = 8$ for the experiments on the ModelNet10 dataset, and $N_{\sf r} = 6$ for these on the CIFAR-10 dataset.
% unless specified otherwise.
We set the transmission bandwidth as $B=10$ kHz, corresponding to the limited bandwidth at the wireless edge.
%The transmission bandwidth between the devices and the edge server is $B=10$ kHz, corresponding to the limited bandwidth at the wireless edge.
The noise power density at the server is $N_0 = -170$ dBm/Hz.
We assume the same distance $d = 240$ m from each edge device to the server, and the path loss is set as $32.6 + 36.7 \lg d$ dB.
The channel between each edge device and the server is independent and modeled as Rician fading, where the Rician factor $\kappa = 1$.
The power budget at each device is set equally as $P_k = P_0$, $k \in \mathcal{K}$.
In addition, the lossy coding precision for the MCR$^2$ based precoding optimization is set to $\epsilon = 10^{-3}$.
We summarize the default values of system parameters in Table \ref{system-parameters}.
The implementation details of the curves in each figure are presented in Table \ref{baselines}.\color{black}
%We set the noise variance $\delta_0^2 = 80$ dBm.
%In addition, the lossy coding precision for the MCR$^2$ based precoding optimization is set to $\epsilon = 10^{-3}$ unless specified otherwise.
\begin{table}[t]
	%	\vspace{-1.5em}
	\caption{Default Values of Simulation Parameters}
	\label{system-parameters}
	\centering
	%	\vspace{-1em}
	%		\scalebox{.9}{
		\begin{tabular}{c|c|c}
			\hline%\hline
			Parameters & CIFAR-10  & ModelNet10  \\
			\hline
			Number of devices & $K=2$ & $K=3$ \\
			\hline
			Feature dimension per device & \multicolumn{2}{c}{$D_k = 8$} \\
			\hline
			Number of transmit antennas per device & \multicolumn{2}{c}{$N_{{\sf t}, k} = 4$} \\
			\hline
			Number of receive antennas & \multicolumn{2}{c}{$N_{\sf r} = 8$} \\
			\hline
			Transmission bandwidth & \multicolumn{2}{c}{$B = 10$ kHz} \\
			\hline
			Noise power density & \multicolumn{2}{c}{$N_0 = -170$ dBm/Hz} \\
			\hline
			Device-server distance & \multicolumn{2}{c}{$d = 240$ m} \\
			\hline
			Path loss &  \multicolumn{2}{c}{$32.6 + 36.7 \lg d$ dB} \\
			\hline
			Rician factor & \multicolumn{2}{c}{$\kappa = 1$} \\
			\hline
			Coding precision for MCR$^2$ precoding & \multicolumn{2}{c}{$\epsilon = 10^{-3}$} \\
			\hline
		\end{tabular}
		%	}
	%		\vspace{-1.5em}
\end{table}

\begin{table*}[t]
	%	\vspace{-1.5em}
	\caption{Implementation Details of the Baselines in Each Figure}
	\label{baselines}
	\centering
	\begin{tabular}{c|c|c|c|c|c}
		\hline%\hline
		Figures & Schemes & Feature encoding & Precoding & Signal detection & Classification  \\
		\hline
		\multirow{3}{*}{Fig. \ref{Fig1}}& Proposed MCR$^2$ & \multirow{3}{*}{MCR$^2$} & MCR$^2$ & None  & MAP  \\
		\cline{2-2}\cline{4-6}
		& LMMSE TRx with NAC \cite{zhijin2021multiuser} & &  LMMSE & LMMSE & CE (noise-aware MLP) \\
%		\cline{2-2}\cline{4-6}
%		&\color{blue} MMSE NN TRx with NAC & & \color{blue} MMSE (NN) & \color{blue} MMSE (NN) & \color{blue} CE (noise-aware MLP) \\
		\cline{2-2}\cline{4-6}
		& Error-free bound & & \multicolumn{2}{c|}{Error-free transmission} & MAP \\
		\hline
		\multirow{3}{*}{Fig. \ref{Fig_DIB_Comparison}} & Proposed MCR$^2$ & MCR$^2$ & MCR$^2$ & None  & MAP  \\
		\cline{2-6}
		& VDDIB \cite{jiawei2023multidevice} & VDDIB & \multicolumn{2}{c|}{Capacity-achieving digital transmission} & VDDIB (MLP) \\
		\cline{2-6}
		& Error-free bound & MCR$^2$ & \multicolumn{2}{c|}{Error-free transmission} & MAP \\
		\hline
		\multirow{5}{*}{Figs. \ref{precoders-ModelNet10} and \ref{precoders-CIFAR-10}} & MCR$^2$ precoder (proposed) & \multirow{5}{*}{MCR$^2$} & MCR$^2$ & \multirow{4}{*}{\makecell{None in subfigures (a); \\ LMMSE in subfigures (b)}}  & \multirow{4}{*}{\makecell{MAP in subfigures (a); \\ None in subfigures (b)}}  \\
		\cline{2-2}\cline{4-4}
		& LMMSE precoder & & LMMSE &   &   \\
		\cline{2-2}\cline{4-4}
		& ITF precoder \cite{iterative_WF} & & IWF &  &  \\
		\cline{2-2}\cline{4-4}
		& Random precoder & &  Random &  &  \\
		\cline{2-2}\cline{4-6}
		& Error-free bound & & \multicolumn{2}{c|}{Error-free transmission} & MAP \\
		\hline
%		\multirow{3}{*}{Figs. \ref{detectors-ModelNet10} and \ref{detectors-CIFAR-10}} & \multirow{3}{*}{MCR$^2$} & \multirow{3}{*}{\makecell{MCR$^2$}} & \multirow{2}{*}{None}  & MAP  \\
%		\cline{5-5}
%		&  &  &  & Nearest subspace classifier  \\
%		\cline{4-5}
%		&  & & LMMSE  & 3-layer MLP \\
%		\hline
	\end{tabular}
\end{table*}
% \ref{Fig2a_ModelNet} and \ref{Fig2a_CIFAR}

\subsubsection{Convergence of Precoding Optimization and Monotonicity of MCR$^2$ Metric}
\begin{figure}
	[t]
	\centering
	\subfigure[$\Delta R$ vs. iteration]
	{\includegraphics[width=.9\columnwidth]{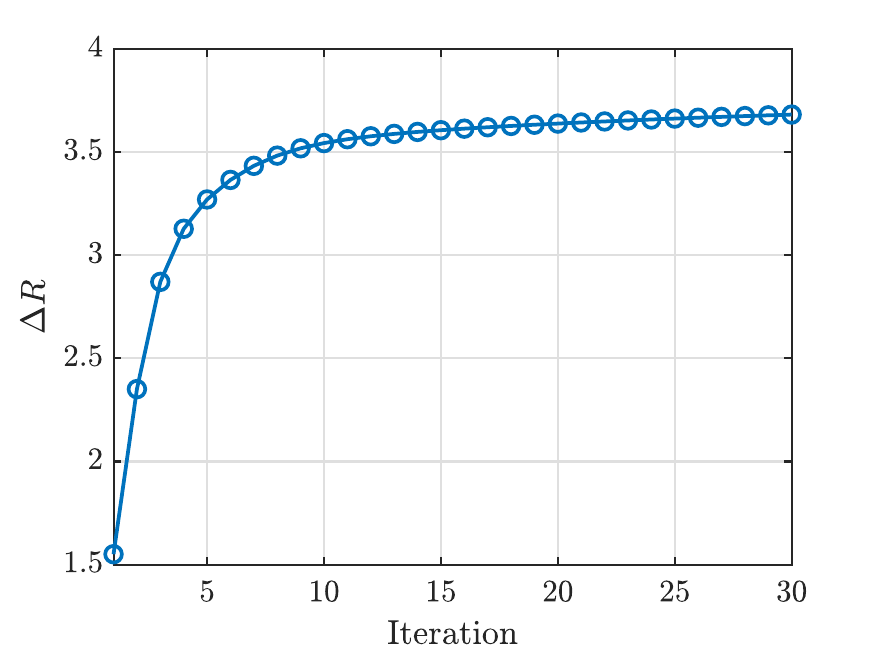}\label{monotonicity_a}} 
	\subfigure[Inference accuracy vs. $\Delta R$]
	{\includegraphics[width=.9\columnwidth]{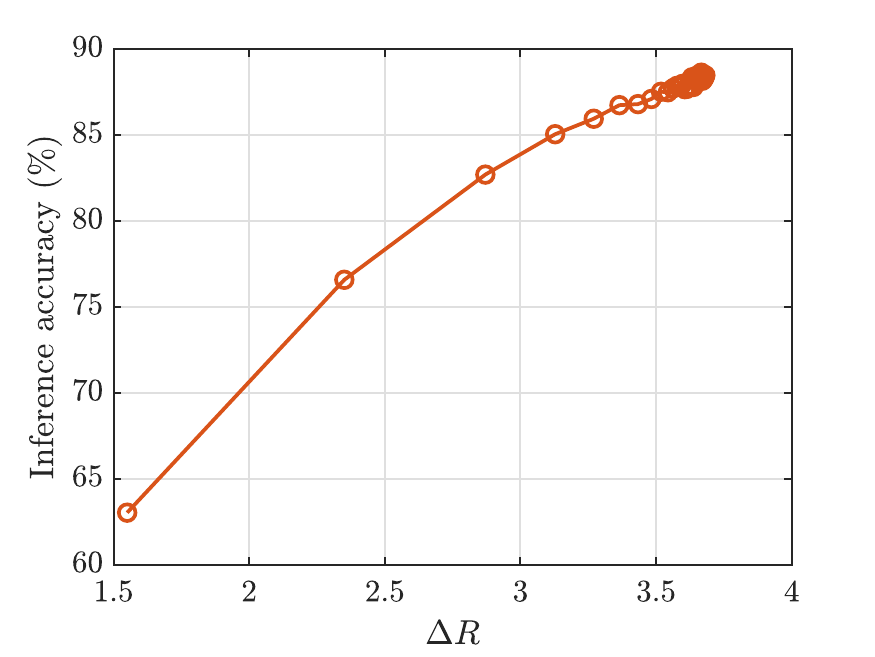}\label{monotonicity_b}}
	\caption{Convergence and monotonicity behaviors of $\Delta R$ on the CIFAR-10 dataset, where $P_0 = 0$ dBm and $T=1$.}
	\label{monotonicity}
\end{figure}

Fig. \ref{monotonicity_a} shows the convergence behavior of the proposed BCA algorithm on the two-view CIFAR-10 dataset. 
The algorithm almost converges within $20$ iterations.
Then, we apply the precoders obtained in each iteration of Fig. \ref{monotonicity_a} to transmit the testing features, where the MAP classifier is utilized to test the inference accuracy.
The results are shown in Fig. \ref{monotonicity_b}, where we plot the inference accuracy versus $\Delta R$ for the $30$ precoders. % of the algorithm % in Fig. \ref{monotonicity_a}
It is observed that the inference accuracy nearly monotonically increases with the increase of $\Delta R$, implying that the MCR$^2$ objective serves an appropriate surrogate measure of the intractable inference accuracy.
%\begin{figure}
%	[t]
%	\centering
%	\subfigure[$\Delta R$ vs. iteration]
%	{\includegraphics[width=.4\columnwidth]{Fig8a_convergence.pdf}\label{monotonicity_a}}~
%	\subfigure[Inference accuracy vs. $\Delta R$]
%	{\includegraphics[width=.4\columnwidth]{Fig8b_monotonicity.pdf}\label{monotonicity_b}}
%	\caption{Convergence and monotonicity behaviors of $\Delta R$ on the CIFAR-10 dataset, where $P_0 = 15$ dBm and $T=1$.}
%	\label{monotonicity}
%\end{figure}

%\begin{figure}
%	[t]
%	\centering
%	\subfigure[$\Delta R$ vs. iteration]
%	{\includegraphics[width=.8\columnwidth]{Revision_Monotonicity_1.pdf}\label{monotonicity_a}} 
%	\subfigure[Inference accuracy vs. $\Delta R$]
%	{\includegraphics[width=.8\columnwidth]{Revision_Monotonicity_2.pdf}\label{monotonicity_b}}
%	\caption{Convergence and monotonicity behaviors of $\Delta R$ on the CIFAR-10 dataset, where $P_0 = 15$ dBm and $T=1$.}
%	\label{monotonicity}
%\end{figure}

%\begin{figure*}
%	[t]
%	\centering
%	\subfigure[$T=1$]
%	{\includegraphics[width=.67\columnwidth]{mcr2_vs_lmmse_T1_Minor.eps}\label{Fig1a_ModelNet}}
%	\subfigure[$T=2$]
%	{\includegraphics[width=.67\columnwidth]{mcr2_vs_lmmse_T2_Minor.eps}\label{Fig1b_ModelNet}}
%	\subfigure[$T=3$]
%	{\includegraphics[width=.67\columnwidth]{mcr2_vs_lmmse_T3_Minor.eps}\label{Fig1c_ModelNet}}
%	\caption{Inference accuracy vs. the maximum transmit power $P_0$ on the ModelNet10 dataset for $T\in \left\{1, 2, 3\right\}$.}
%	\label{Fig1}
%\end{figure*}

\begin{figure*}
	[t]
	\centering
	\subfigure[$T=1$]
	{\includegraphics[width=.67\columnwidth]{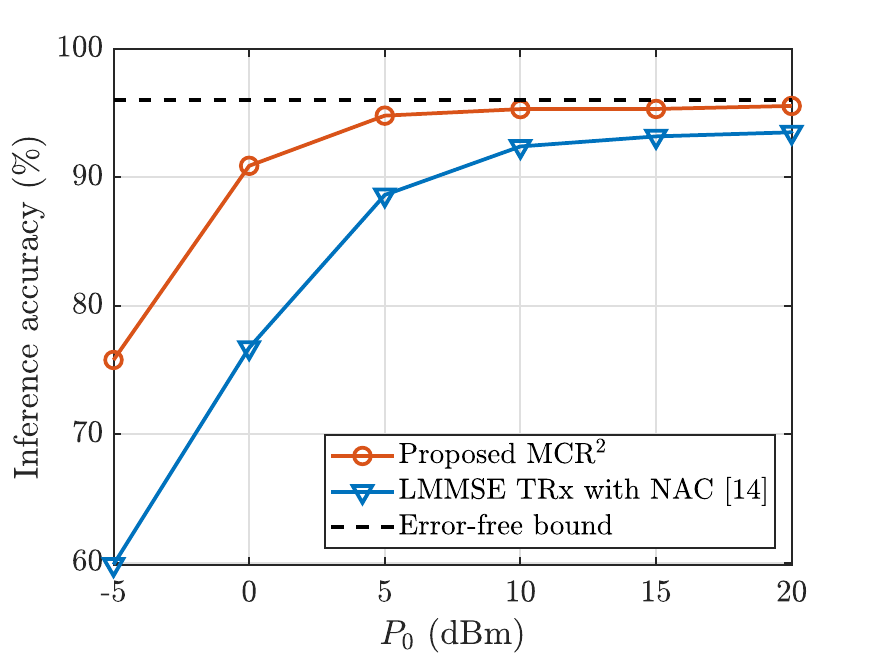}\label{Fig1a_ModelNet}}
	\subfigure[$T=2$]
	{\includegraphics[width=.67\columnwidth]{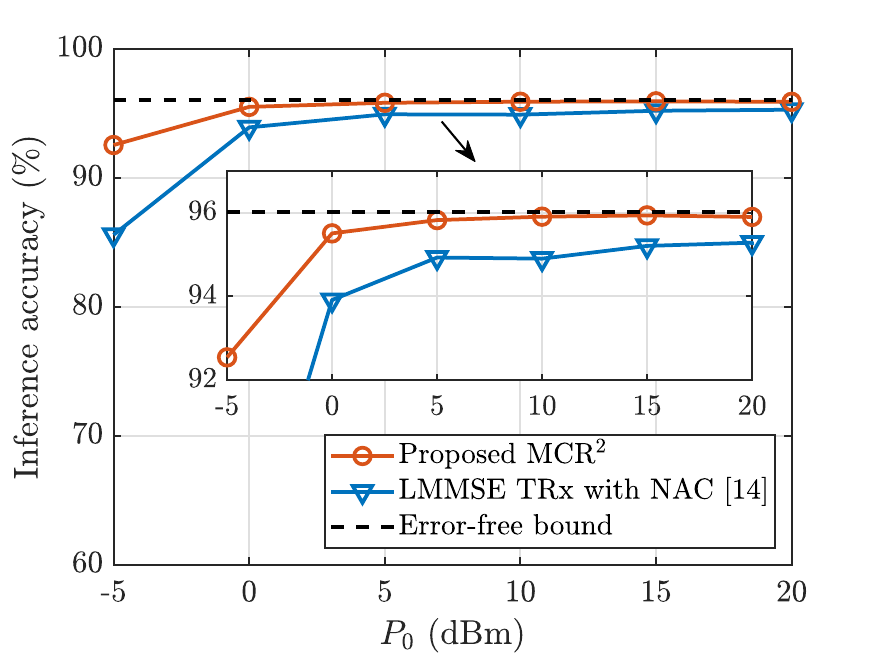}\label{Fig1b_ModelNet}}
	\subfigure[$T=3$]
	{\includegraphics[width=.67\columnwidth]{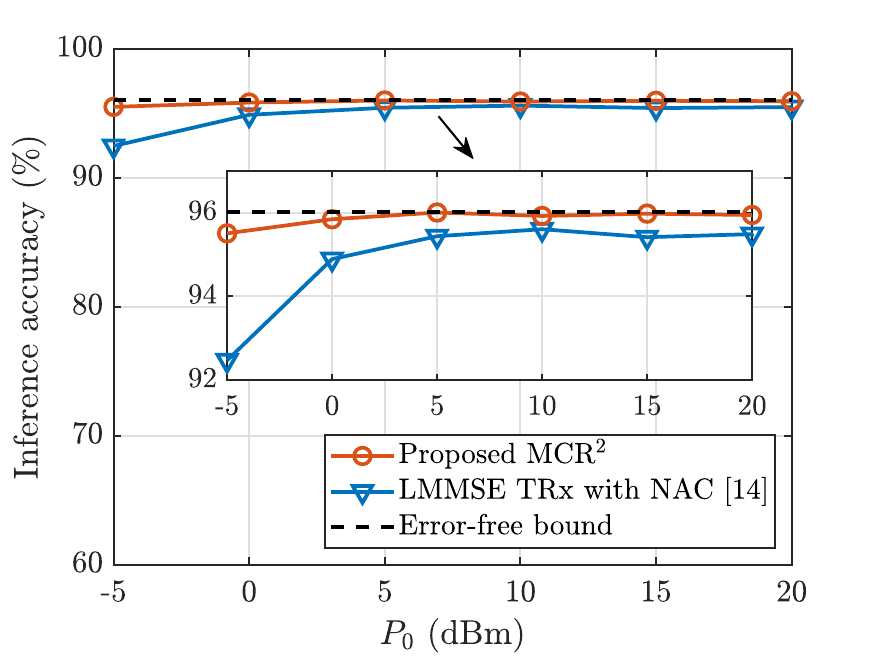}\label{Fig1c_ModelNet}}
	\caption{Inference accuracy vs. the maximum transmit power $P_0$ on the ModelNet10 dataset for $T\in \left\{1, 2, 3\right\}$.}
	\label{Fig1}
\end{figure*}

\subsubsection{Comparisons with Existing Task-Oriented Communication Methods}
We adopt the method in \cite{zhijin2021multiuser} as a baseline with necessary modifications made to accommodate the considered inference task and the multi-antenna setting.
Specifically, the baseline adopts the same MCR$^2$ feature encoding network as in the proposed method, but applies the LMMSE precoder\footnote{The implementation details of the LMMSE precoder are presented in Appendix \ref{LMMSE_precoder}.} at the edge devices and the LMMSE detector at the server, resulting in an inconsistent learning-communication design objective.
The classification network is a three-layer multilayer perceptron (MLP) trained with the recovered features instead of the noise-free ones to guarantee robustness.
%For different transmit power and transmit time slot, 
%We train dedicated classification networks for different transmit power (SNR) and time-slot budgets to secure the best baseline performance.
We train multiple individual classification networks, each specialized for a different transmit power and transmit time-slot budget.
We refer to this baseline as the LMMSE transceiver (TRx) with noise-aware classifier (NAC), sometimes also abbreviated as LMMSE.

%\color{blue}
%Moreover, since the LMMSE based design is not optimal in terms of MSE for non-Gaussian source, we implement another baseline with NN based transceivers that directly minimize the MSE loss.
%The recovered features are then fed into the NAC for classification.
%We refer to this baseline as the MMSE NN TRx with NAC.
%%The classifier is also noise-adaptive, which follows the same design as in the 
%\color{black}

Fig. \ref{Fig1} compares the inference accuracy of the proposed method with the LMMSE benchmark on the ModelNet10 dataset under different transmit time-slot budgets.
%We train a total of $18$ classification networks for the LMMSE benchmark, each of which is dedicated to a different transmit power and transmit time slot.
%During the training stage, 
% using the recovered features by LMMSE detection instead of the noise-free features.
%We compare the inference accuracy of the proposed method with the LMMSE benchmark on the ModelNet10 dataset, as shown in Fig. \ref{Fig1}.
%The LMMSE benchmark applies the LMMSE precoder at the edge devices and the LMMSE detector at the server,
%and then feeds the recovered features into a three-layer multilayer perceptron (MLP) to obtain the final classification result.
For the case that a single time slot is used to transmit one $\dot{\mbf{z}}$, there exists a significant gap on the inference accuracy between the proposed method and the LMMSE benchmark.
The gap diminishes but does not disappear in the high SNR regime.
This is because $\mbf{H} \mbf{V}$ is a fat matrix ($TN_{\sf r} < \frac{D}{2}$) when $T=1$, so that the LMMSE detector fails.
When two or three time slots are used for transmission, although $\mbf{H}\mbf{V}$ becomes a tall matrix, we can still observe a noticeable performance gain at low and intermediate SNRs.
The LMMSE benchmark finally achieves comparable accuracy with our design at high SNR.
The above observations validate the superior performance of the proposed design, especially when the communication resources are insufficient.
%The same comparisons are also made on the CIFAR-10 dataset in Fig. \ref{Fig1-CIFAR}.

\begin{figure*}[t]
	\centering
	%	\vspace{-2em}
	\subfigure[Proposed method vs. VDDIB digital (FDMA)]
	{\includegraphics[width=.9\columnwidth]{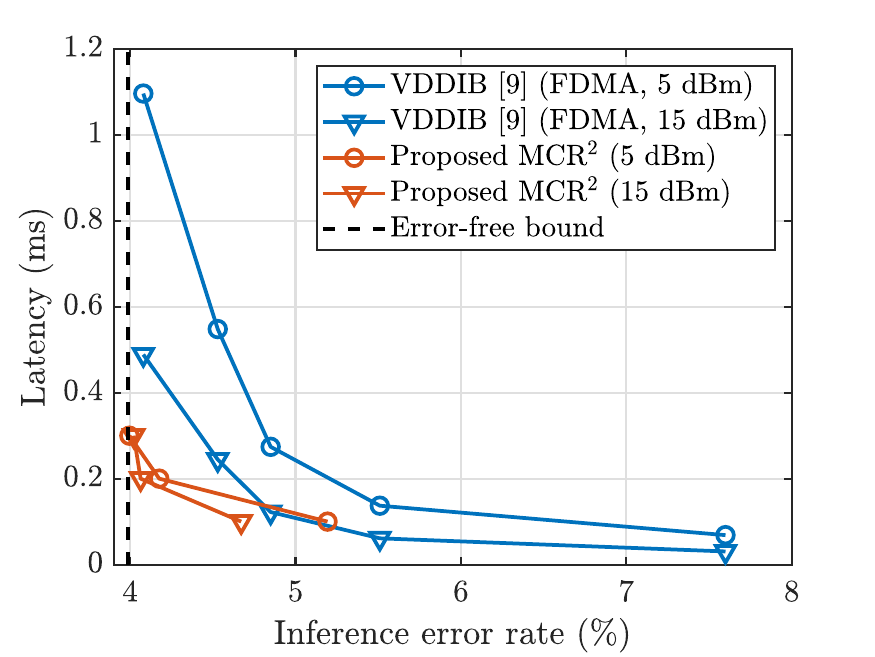}\label{Fig_DIB_FDMA}} \hfil
	\subfigure[Proposed method vs. VDDIB digital (TDMA)]
	{\includegraphics[width=.9\columnwidth]{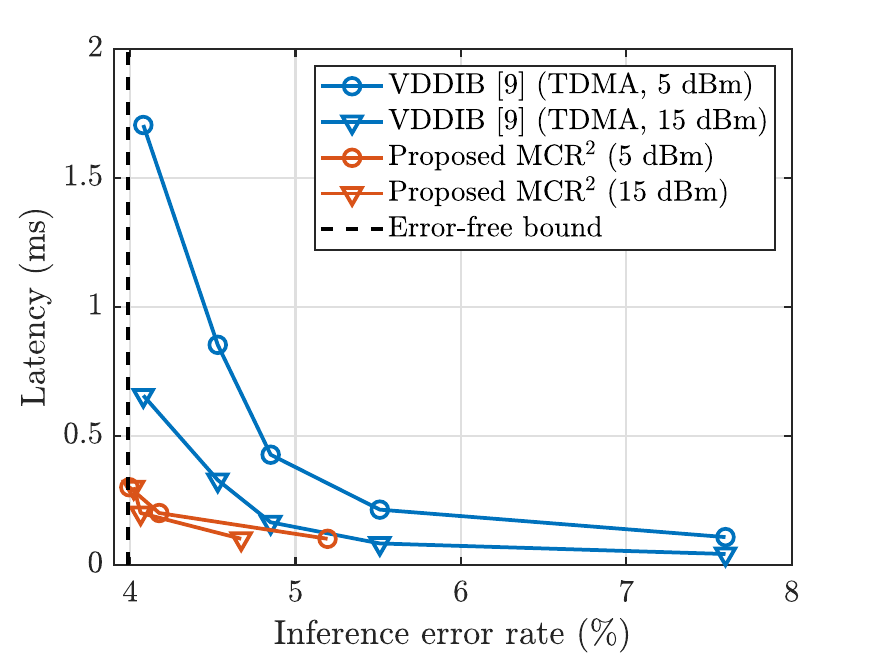}\label{Fig_DIB_TDMA}}
	\caption{The tradeoff between communication latency and inference error rate on the ModelNet10 dataset.
		%		We vary $T \in \left\{1, 2, 3\right\}$ and $D_{{\sf VDDIB},k} \in \left\{2, 4, 8, 16, 32\right\}$ to obtain the curves of the proposed method and the VDDIB benchmark, respectively.
		We provide the curves at both $P_0 = 5$ dBm and $15$ dBm.
		The VDDIB benchmark adopts the same VGG11 backbone for feature encoding as in the proposed method.}
	\label{Fig_DIB_Comparison}
	%	\vspace{-1.5em}
\end{figure*}

Consider the tradeoff between communication latency and classification accuracy.
The communication latency of the proposed method is $L_{\sf proposed} = \frac{T}{B}$, which is irreverent to the power budget and channel condition.
We choose the variational distributed deterministic information bottleneck (VDDIB) method \cite{jiawei2023multidevice} for comparison. % comparing the latency-accuracy tradeoff. % digital transmission 
The VDDIB benchmark, built on the information bottleneck principle \cite{ib2000, dib2021}, is a deep learning method for task-aware lossy compression and classification.
%It restricts each feature dimension to be a binary bit.
The feature encoding and classification networks are jointly trained by the VDDIB objective \cite{jiawei2023multidevice}.
We adopt the same feature encoding backbone as in the proposed method, and adopt a three-layer MLP for classification.
The VDDIB benchmark restricts each feature dimension to be a binary bit, which is naturally compatible with digital communications.
Following \cite{jiawei2023multidevice}, we assume capacity-achieving digital transmission of features in this benchmark. %, so that the latency can be calculated based on the channel capacity.
Notice that this induces a misalignment between the learning and communication objectives since digital communication aims at the accurate recovery of transmitted feature bits. %, which can cause significant delays.
The accurate recovery can often be at the expense of significant delays.
%The received features are then feed into the classifier built upon a three-layer multilayer perceptron (MLP).
%To avoid multi-user interference, 
%In the VDDIB benchmark, w
In the VDDIB benchmark, both frequency-division multiple access (FDMA) and time-division multiple access (TDMA) are considered among the devices. % for feature transmission.
For FDMA, each device is assigned with $\frac{B}{K}$ bandwidth, and the corresponding communication latency is given by $L_{\sf VDDIB, FDMA} = \max_{k \in \mathcal{K}} \left\{\frac{D_{{\sf VDDIB},k}}{C_{{\sf FDMA}, k}} \right\}$, where $D_{{\sf VDDIB},k}$ is the number of binary bits to be transmitted at device $k$, and $C_{{\sf FDMA}, k} = \frac{B}{K} \log \det \left(\mbf{I} + \frac{\mbf{H}_k \mbf{Q}^\star_k \mbf{H}_k^{\sf H}}{N_0 \frac{B}{K}} \right)$ with $\mbf{Q}^\star_k$ being the optimal transmit covariance matrix given by the eigenmode transmission \cite{tse2005fundamentals}.
For TDMA, the communication latency is given by $L_{\sf VDDIB, TDMA} = \sum_{k \in \mathcal{K}} \frac{D_{{\sf VDDIB},k}}{C_{{\sf TDMA}, k}}$, where $C_{{\sf TDMA}, k} = B \log \det \left(\mbf{I} + \frac{\mbf{H}_k \mbf{Q}^\star_k \mbf{H}_k^{\sf H}}{N_0 B} \right)$.
%The cross-time-slot precoding framework developed in Section \ref{subsec_CM} allows us to achieve a different latency-accuracy tradeoff by adjusting $T$.
%Owing to the cross-time-slot precoding framework developed in Section \ref{subsec_CM}, we can simply adjust $T$ to achieve a different latency-accuracy tradeoff.

In Fig. \ref{Fig_DIB_Comparison}, we vary $T \in \left\{1, 2, 3\right\}$ and $D_{{\sf VDDIB},k} \in \left\{2, 4, 8, 16, 32\right\}$ to obtain the tradeoff curves of the proposed method and the VDDIB benchmark, respectively.
For the proposed method, by comparing the curves at $P_0 = 5$ dBm and $15$ dBm, we observe that increasing the transmit power improves the inference accuracy while maintaining the same latency.
For the VDDIB benchmark, in contrast, increasing the transmit power reduces the latency (by improving the channel capacity) while maintaining the same accuracy (since the number of transmitted bits is fixed).
Compared to the VDDIB benchmark,
our proposed method achieves a better latency-accuracy tradeoff, i.e, with a given latency requirement, a lower inference error rate is achieved, and vice versa. %It is shown that
This is because the proposed method does not rely on the accurate recovery of transmitted features, but only requires the different classes of received features to be maximally separated, which aligns well with the objective of the classification task.
%This communication objective aligns well with the objective of a classification task.
In other words, we tolerate any kind of feature distortion, such as translation, rotation, scaling, noise, or even compression, as long as it does not deteriorate the classification accuracy.
This tolerance/relaxation can avoid unnecessary communication costs for feature recovery, reflected in Fig. \ref{Fig_DIB_Comparison} as the reduced latency of the proposed method.
%In other words, we tolerate intra-class distortion, rotation, or even compression since it does not deteriorate the classification accuracy.
%In other words, we tolerate the distortion occurred intra-class since it does not deteriorate the classification accuracy. % that occurs intra-class
%This tolerance can be regarded as a relaxation on the communication requirement, which avoids the unnecessary communication costs for feature recovery, reflected in Fig. \ref{Fig_DIB_Comparison} as the reduced latency of the proposed method. % for feature recovery
% targeting the same classification accuracy.
%whose latency is restricted by the channel capacity.
%In other words, we tolerate the distortion that occurs intra-class since it does not deteriorate the separability of different classes, and hence the classification accuracy.
%On the other hand, accurate feature reconstruction in VDDIB is indeed a more stringent communication requirement, whose latency is restricted by the channel capacity.
Moreover, we also observe that the latency advantage is especially large in the high inference accuracy regime, located on the left side in Figs. \ref{Fig_DIB_FDMA} and \ref{Fig_DIB_TDMA}.
%which is favorable in real deployment.
%which is favorable for low-latency and high-accuracy inference in real deployment.

%Fig. \ref{Fig_DIB_Comparison} compares the tradeoff curve of the proposed method against that of the VDDIB benchmark with FDMA and TDMA, respectively.

%The VDDIB benchmark is a deep learning method for task-aware lossy compression, which trains the feature encoders to output binary bits based on some variants of the information bottleneck principle \cite{ib2000, dib2021}.

%bottleneck is restricted to digital; jointly trained

%The relationship between the achieved classification accuracy and the number of transmitted bits is shown in Fig. 4.

%\begin{figure}
%	[t]
%	\centering
%	\subfigure[Inference accuracy vs. $N_{\sf r}$ by MAP classification]
%	{\includegraphics[width=.4\columnwidth]{Fig11a_acc_modelnet.pdf}\label{Fig2a_ModelNet}}~
%	\subfigure[MSE vs. $N_{\sf r}$ by LMMSE detection]
%	{\includegraphics[width=.4\columnwidth]{Fig11b_mse_modelnet.pdf}\label{Fig2b_ModelNet}}
%	\caption{Inference accuracy and MSE performance for different precoders on the ModelNet10 dataset, where $P_0 = 10$ dBm and $T=2$.}
%	\label{precoders-ModelNet10}
%\end{figure}
\begin{figure*}
	[t]
	\centering
	\subfigure[Inference accuracy vs. $N_{\sf r}$ by MAP classification]
	{\includegraphics[width=.9\columnwidth]{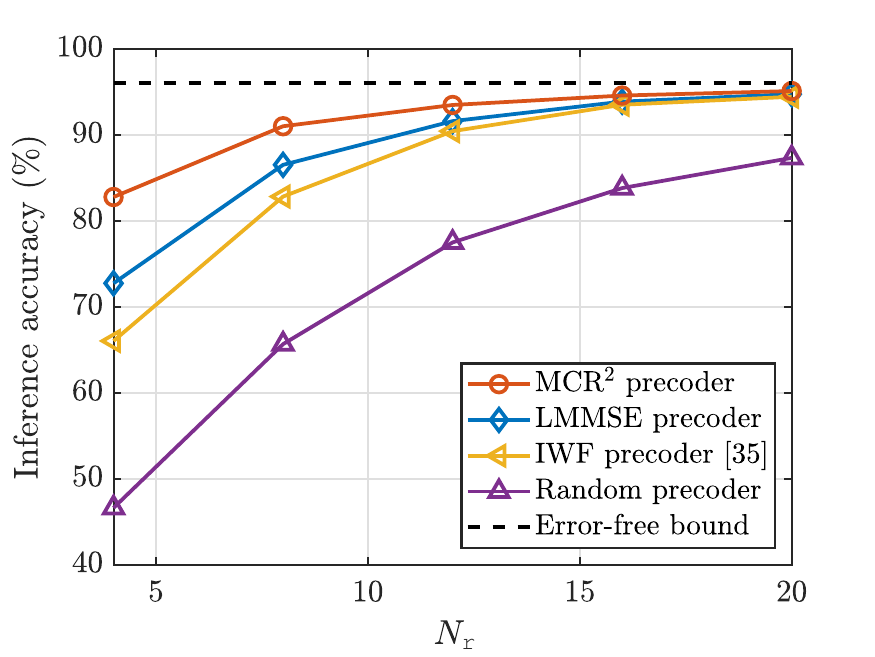}\label{Fig2a_ModelNet}} \hfil
	\subfigure[MSE vs. $N_{\sf r}$ by LMMSE detection]
	{\includegraphics[width=.9\columnwidth]{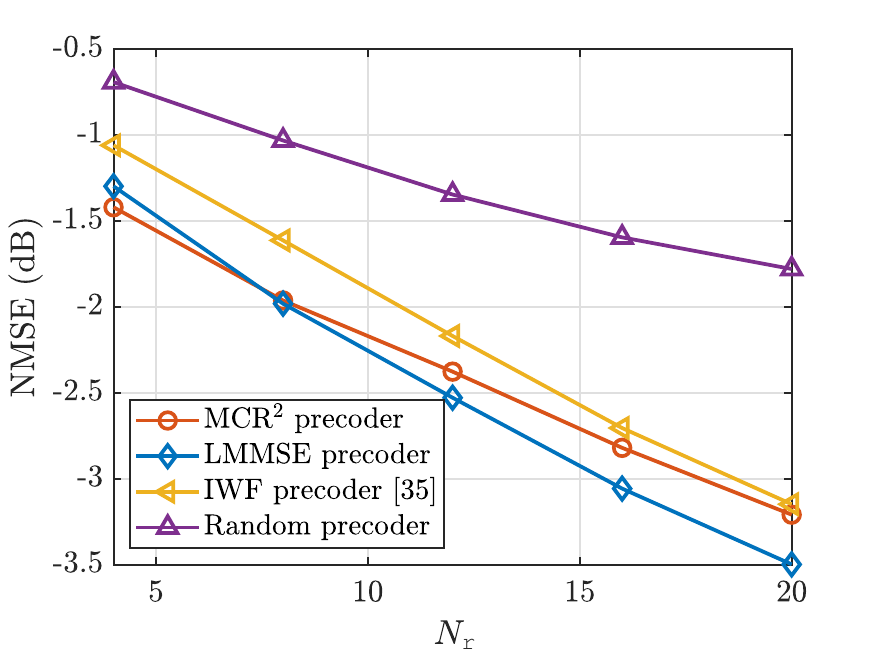}\label{Fig2b_ModelNet}}
	\caption{Inference accuracy and MSE performance for different precoders on the ModelNet10 dataset, where $P_0 = 0$ dBm and $T=1$.}
	\label{precoders-ModelNet10}
\end{figure*}

\begin{figure*}
	[t]
	\centering
	\subfigure[Inference accuracy vs. $N_{\sf r}$ by MAP classification]
	{\includegraphics[width=.9\columnwidth]{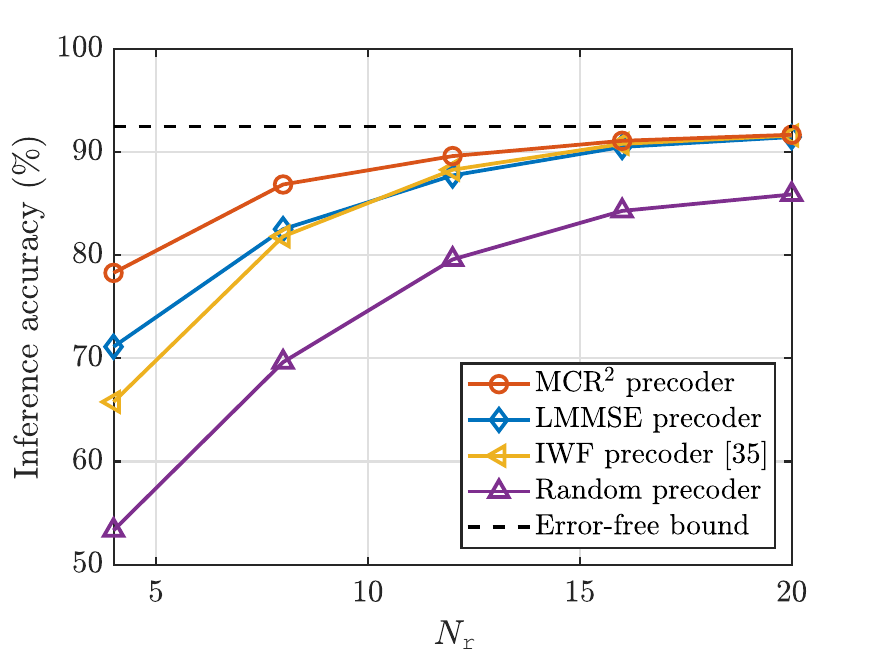}\label{Fig2a_CIFAR}} \hfil
	\subfigure[MSE vs. $N_{\sf r}$ by LMMSE detection]
	{\includegraphics[width=.9\columnwidth]{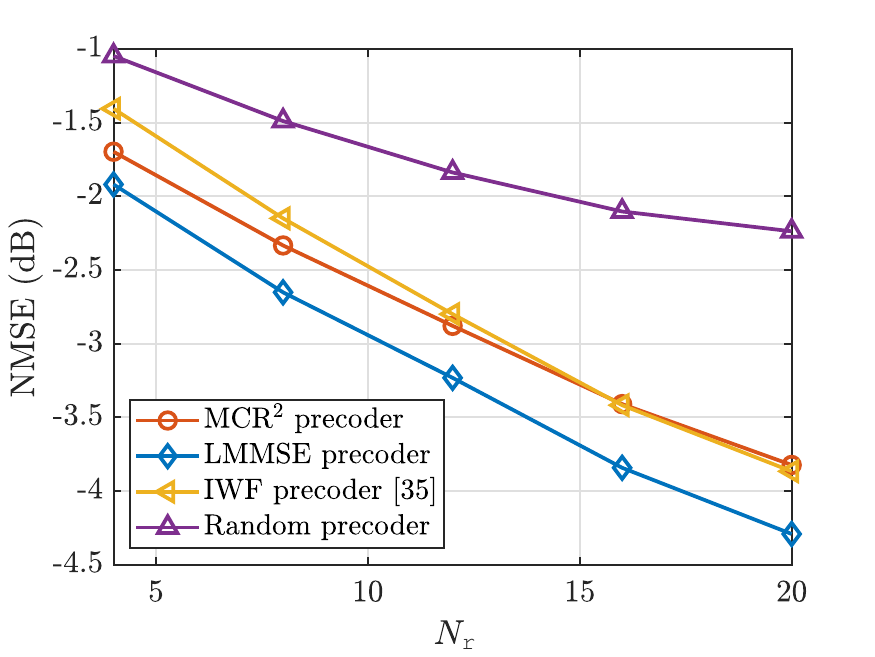}\label{Fig2b_CIFAR}}
	\caption{Inference accuracy and MSE performance for different precoders on the CIFAR-10 dataset, where $P_0 = 0$ dBm and $T=1$.}
	\label{precoders-CIFAR-10}
\end{figure*}

\subsubsection{Comparisons on Different Precoders}
To highlight the necessity of the MCR$^2$ based precoding design, we compare various precoding schemes on the ModelNet10 dataset in Fig. \ref{precoders-ModelNet10}.
Particularly, all the benchmarks adopt the same MCR$^2$ feature encoding network, followed by different precoding schemes.
The LMMSE precoder is described in Appendix \ref{LMMSE_precoder}.
The iterative water-filling (IWF) precoder \cite{iterative_WF} is a capacity-achieving precoder for the Gaussian multiple-access channel.
We test the inference accuracy of each scheme by applying MAP classification to the received features, as shown in Fig. \ref{Fig2a_ModelNet}.
The data recovery performance in terms of the normalized MSE (NMSE) is presented in Fig. \ref{Fig2b_ModelNet} by applying LMMSE detection.
It is observed that the MCR$^2$ precoder achieves the highest inference accuracy, but does not necessarily have the lowest MSE.
This is because the MCR$^2$ objective does not force any sample-wise consistency between the transmitted and recovered features.
Instead, it only promotes class-wise separability on the received features regardless of the MSE distortion.
%This implies that t
The MCR$^2$ precoder can even behave as a compressor to compress the transmitted features, as long as different classes are separated (this phenomenon is validated in Fig. \ref{ellipsoid_lp}).
However, relying on the received compressed version for feature recovery is indeed a challenging task.
%However, recovering the transmitted features based on the received compressed version is indeed a challenging task.
This explains why the MCR$^2$ precoder may exhibit a worse MSE performance compared to the LMMSE precoder.
% so that the LMMSE detector can hardly recover the transmitted features.
%Therefore, the proposed method exhibits a worse MSE performance compared to the LMMSE precoder.
%This implies that the MSE is not a good metric to serve the inference task.
%Moreover, applying the precoding design targeting at MSE minimization or throughput maximization is not optimal when 
The results obtained on the CIFAR-10 dataset is consistent with the above observations, as presented in Fig. \ref{precoders-CIFAR-10}.

\section{Concluding Remarks}\label{Section_Conclusion}
In this paper, we studied the design of a multi-device edge inference system over a MIMO multiple-access channel.
We separated the design process into two parts: the learning task (i.e., feature encoding and classification) and physical-layer communication (i.e., precoding), while maintaining a consistent design objective for inference accuracy maximization.
We invoked the MCR$^2$ objective as a surrogate measure of the intractable inference accuracy to formulate the precoding optimization problem.
Moreover, the MCR$^2$ objective also served as the loss function to train the feature encoders, enabling a channel-adaptive MAP classifier to infer the classification result at the edge server.
Our simulation results highlighted the superior performance of our proposed method.
It outperformed the LMMSE benchmark and various baselines, demonstrating its effectiveness and potential for practical application.

From a more general perspective, this work serves as the first attempt towards the synergistic alignment of learning and communication objectives in task-oriented communications by modularized design.
We emphasize that the classification task considered in this work is simple in the sense that, the edge server generates the inference result based on the feature distribution without the need of a downstream NN.
% signal detection to feed the recovered features into
However, for general tasks such as speech recognition, machine translation, and video caption, 
%For more complicated generative tasks,
a server-side downstream NN is essential to generate these multifarious contents.
This calls for a line of research on signal detection design to recover transmitted features in a task-oriented manner.
%anticipate
%As for the signal detection objective, solely relying on

\appendices
\section{Implementation Details of the LMMSE Precoder} \label{LMMSE_precoder}
The LMMSE detector is given by
\begin{align} \label{LMMSE_det}
	\hat{\dot{\mbf{z}}} = \mbf{\Sigma} \mbf{V}^{\sf H} \mbf{H}^{\sf H} \left(\mbf{H} \mbf{V} \mbf{\Sigma}\mbf{V}^{\sf H}\mbf{H}^{\sf H} + \delta_0^2 \mbf{I}\right)^{-1} \mbf{y}.
\end{align}
Based on \eqref{LMMSE_det}, the MSE of the LMMSE detector can be derived as
\begin{align} \label{LMMSE}
	\mathrm{MSE} % &= \mathbb{E} \left[\big\|\hat{\dot{\mbf{z}}} - \dot{\mbf{z}}\big\|_2^2\right] \nonumber \\
	= \mathrm{tr} \left(\mbf{\Sigma} - \mbf{\Sigma} \mbf{V}^{\sf H} \mbf{H}^{\sf H} \left(\mbf{H} \mbf{V} \mbf{\Sigma}\mbf{V}^{\sf H}\mbf{H}^{\sf H} + \delta_0^2 \mbf{I}\right)^{-1} \mbf{H} \mbf{V} \mbf{\Sigma} \right).
\end{align}
The LMMSE precoder aims to minimize \eqref{LMMSE}, subject to the transmit power constraint at each device (by ignoring constant terms):
\begin{align}
	\min_{\left\{\mbf{V}_k \in \mathcal{V}_k\right\}_{k\in\mathcal{K}}} -\mathrm{tr} \left( \mbf{\Sigma} \mbf{V}^{\sf H} \mbf{H}^{\sf H} \left(\mbf{H} \mbf{V} \mbf{\Sigma}\mbf{V}^{\sf H}\mbf{H}^{\sf H} + \delta_0^2 \mbf{I}\right)^{-1} \mbf{H} \mbf{V} \mbf{\Sigma} \right).
\end{align}
%\begin{subequations}
%	\begin{align}
%		\min_{\left\{\mbf{V}_k\right\}_{k\in\mathcal{K}}}~ & \mathrm{tr} \left(\mbf{\Sigma} - \mbf{\Sigma} \mbf{V}^{\sf H} \mbf{H}^{\sf H} \left(\mbf{H} \mbf{V} \mbf{\Sigma}\mbf{V}^{\sf H}\mbf{H}^{\sf H} + \delta_0^2 \mbf{I}\right)^{-1} \mbf{H} \mbf{V} \mbf{\Sigma} \right)  \\ 
%		\operatorname{ s.t. } ~~~
%		& \frac{1}{T} \mathrm{tr} \left(\mbf{V}_k\mbf{\Sigma}^{(kk)} \mbf{V}_k^{\sf H}\right) \leq P_k, ~~k \in \mathcal{K}.
%	\end{align}
%\end{subequations}
The above problem is a fractional programming problem.
We apply the quadratic transform \cite[Theorem 2]{shen2018fractional} to equivalently transform the above problem as
\begin{align} \label{quadratic_program}
	\min_{\left\{\mbf{V}_k \in \mathcal{V}_k\right\}_{k\in \mathcal{K}}, \mbf{R}} ~ & -2 \mathrm{Re} \left\{ \mathrm{tr} \left(\mbf{R}^{\sf H} \mbf{H} \mbf{V} \mbf{\Sigma} \right) \right\} \nonumber \\
	&+ \mathrm{tr} \left(\mbf{R}^{\sf H} \left(\mbf{H} \mbf{V} \mbf{\Sigma}\mbf{V}^{\sf H} \mbf{H}^{\sf H} + \delta_0^2 \mbf{I}\right) \mbf{R}\right) ,
\end{align}
%\begin{subequations} \label{quadratic_program}
%	\begin{align}
%		\min_{\left\{\mbf{V}_k\right\}_{k\in \mathcal{K}}, \mbf{R}} ~ & -2 \mathrm{Re} \left\{ \mathrm{tr} \left(\mbf{R}^{\sf H} \mbf{H} \mbf{V} \mbf{\Sigma} \right) \right\} \nonumber \\
%		&+ \mathrm{tr} \left(\mbf{R}^{\sf H} \left(\mbf{H} \mbf{V} \mbf{\Sigma}\mbf{V}^{\sf H} \mbf{H}^{\sf H} + \delta_0^2 \mbf{I}\right) \mbf{R}\right)  \\ 
%		\operatorname{ s.t. } ~~~~
%		& \frac{1}{T} \mathrm{tr} \left(\mbf{V}_k\mbf{\Sigma}^{(kk)} \mbf{V}_k^{\sf H}\right) \leq P_k, ~~k \in \mathcal{K},
%	\end{align}
%\end{subequations}
where $\mbf{R} \in \mathbb{C}^{TN_{\sf r} \times TN_{\sf r}}$ refers to the introduced auxiliary matrix.

To solve the problem in \eqref{quadratic_program}, we propose to alternatingly optimize $\mbf{R}$ and one of the precoders $\mbf{V}_k$ with the others fixed.
For any fixed $\left\{\mbf{V}_k\right\}_{k\in \mathcal{K}}$, the optimization w.r.t. $\mbf{R}$ is convex, and the optimal solution is given by
%the optimal $\mbf{R}$ is given by
\begin{align}
	\mbf{R} = \left(\mbf{H} \mbf{V} \mbf{\Sigma} \mbf{V}^{\sf H} \mbf{H}^{\sf H} + \delta_0^2 \mbf{I}\right)^{-1} \mbf{H} \mbf{V} \mbf{\Sigma}.
\end{align}
Given $\mbf{R}$ and $\left\{\mbf{V}\right\}_{j\in \mathcal{K}\backslash \{k\}}$, the optimization of $\mbf{V}_k$ reduces to the following QCQP:
\begin{subequations}
	\begin{align}
		\min_{\mbf{V}_k} \quad & -2\mathrm{Re} \left\{\mathrm{tr} \left(\mbf{R}^{\sf H} \mbf{H}_k \mbf{V}_k  \mbf{J}_{k}\right) \right\} \nonumber \\
		& +\mathrm{tr} \left(\mbf{R}^{\sf H}  \mbf{H}_k \mbf{V}_k \mbf{\Sigma}^{(kk)} \mbf{V}_k^{\sf H} \mbf{H}_k^{\sf H}  \mbf{R}\right) \label{LMMSE_obj}\\ 
		\operatorname{ s.t. } \quad
		& \mathrm{tr} \left(\mbf{V}_k\mbf{\Sigma}^{(kk)} \mbf{V}_k^{\sf H}\right) \leq T P_k, \label{LMMSE_cons}
	\end{align}
\end{subequations}
where $\mbf{J}_k \triangleq \mbf{\Sigma}^{(k)} - \sum_{q\neq k} \mbf{\Sigma}^{(kq)} \mbf{V}_q^{\sf H} \mbf{H}_q^{\sf H}  \mbf{R} $.
The matrix $\mbf{\Sigma}^{(k)}$ in $\mbf{J}_k$ denotes row $m_k$ to row $n_k$ of $\mbf{\Sigma}$, where $m_k \triangleq \sum_{i=0}^{k-1}\frac{D_i}{2} +1$ and $n_k \triangleq \sum_{i=1}^k \frac{D_i}{2}$, with $D_0 \triangleq 0$;
$\mbf{\Sigma}^{(kq)} \triangleq \mathbb{E} \left[\dot{\mbf{z}}_k \dot{\mbf{z}}_q^{\sf H}\right]-\bsm{\mu}_k \bsm{\mu}_q^{\sf H}$ is the matrix formed by row $m_k$ to row $n_k$ and column $m_q$ to column $n_q$ of $\mbf{\Sigma}$.
%, where $u \triangleq \sum_{i=0}^{q-1}D_i/2 +1$ and $v \triangleq \sum_{i=1}^q D_i/2$.
By exploiting the first-order optimality conditions of the above problem, we arrive at its optimal solution as
\begin{align}
	\mbf{V}_k =& \left(\mbf{H}_k^{\sf H}\mbf{R}\mbf{R}^{\sf H}\mbf{H}_k + \mu_k \mbf{I}\right)^{-1} \mbf{H}_k^{\sf H} \mbf{R} \mbf{J}_k^{\sf H} \left(\mbf{\Sigma}^{(kk)}\right)^{-1},
\end{align}
where $\mu_k$ is the dual variable associated with the power constraint \eqref{LMMSE_cons}.

%\section{Implementation Details of the MMSE NN TRx with NAC Baseline} \label{MMSE_precoder_decoder}

\bibliographystyle{IEEEtran}%By using IEEEtrans, the number can be displayed.
\bibliography{IEEEabrv,mybib}

\end{document}